\documentclass[12pt,a4paper]{article}

\usepackage[english]{babel}

\usepackage[authoryear]{natbib}
\usepackage[utf8]{inputenc}
\usepackage{comment}
\usepackage{algorithmic}
\usepackage{booktabs}
\usepackage{algorithm}
\usepackage[official]{eurosym}
\usepackage{multirow}
\usepackage{amsmath, amsthm, amssymb, amsfonts, graphicx,tabularx,rotate}
\usepackage{arydshln} 

\usepackage{longtable, lscape, adjustbox,empheq}
\usepackage[figuresright]{rotating}
\usepackage{multicol}
\usepackage{caption,rotfloat,subcaption}
\usepackage{afterpage,url}
\usepackage[tmargin=2.5cm,bmargin=2.25cm,lmargin=2.25cm,rmargin=2.25cm]{geometry}
\usepackage{setspace}
\usepackage[dvipsnames]{xcolor}
\usepackage[hidelinks]{hyperref}

\usepackage{placeins}
\makeatletter
\AtBeginDocument{%
  \expandafter\renewcommand\expandafter\subsection\expandafter
    {\expandafter\@fb@secFB\subsection}%
  \newcommand\@fb@secFB{\FloatBarrier
    \gdef\@fb@afterHHook{\@fb@topbarrier \gdef\@fb@afterHHook{}}}%
  \g@addto@macro\@afterheading{\@fb@afterHHook}%
  \gdef\@fb@afterHHook{}%
}
\makeatother
\begin{document}


\setcounter{page}{0}
\pagenumbering{Alph}
\thispagestyle{empty} 

\doublespacing
\begin{center}
\LARGE{\textbf{Forecasting the Volatility of Energy Transition Metals}}\\

\normalsize{\today}\\\vspace{.25cm}

\vspace{0.5cm} 
\Large{Andrea Bastianin$^{a,b}$\qquad Xiao Li$^{a,b,c}$\qquad Luqman Shamsudin$^{b,d}$}\\

\vspace{0.3cm} 
\small{%
$^{a}$ Department of Economics, Management, and Quantitative Methods, University of Milan, Milan, Italy\\
$^{b}$ Fondazione Eni Enrico Mattei, Milan, Italy\\
$^{c}$ Department of Economics, University of Pavia, Pavia, Italy\\
$^{d}$ Department of Environmental Science and Policy, University of Milan, Milan, Italy}\\

\vspace{0.5cm}
\end{center}

\noindent\textbf{Corresponding author}: Xiao Li, Department of Economics, Management, and Quantitative Methods (DEMM). University of Milano. Via Conservatorio, 7. 20122 - Milano - Italy. Email: \url{xiao.li@unimi.it}%

\vspace{.5cm}

\noindent\textbf{Acknowledgments}: Andrea Bastianin and Luqman Shamsudin gratefully acknowledge funding from the Italian Ministry MIUR under the PRIN-PNRR project GREESCO (grant G53D23006810001).

\vspace{.5cm}

\noindent\textbf{Declaration of interests}: The authors declare that they have no known competing financial interests or personal relationships that could have appeared to influence the work reported in this paper.

\newpage

\setcounter{footnote}{0}
\pagenumbering{arabic}
\doublespacing
\normalsize


\setcounter{page}{0}
\pagenumbering{Alph}
\thispagestyle{empty} 

\doublespacing
\begin{center}
\LARGE{\textbf{Forecasting the Volatility of Energy Transition Metals}}\\

\end{center}

\small{\noindent\textbf{Abstract.} The transition to a cleaner energy mix, essential for achieving net-zero greenhouse gas emissions by 2050, will significantly increase demand for metals critical to renewable energy technologies. Energy Transition Metals (ETMs), including copper, lithium, nickel, cobalt, and rare earth elements, are indispensable for renewable energy generation and the electrification of global economies. However, their markets are characterized by high price volatility due to supply concentration, low substitutability, and limited price elasticity. This paper provides a comprehensive analysis of the price volatility of ETMs, a subset of Critical Raw Materials (CRMs). Using a combination of exploratory data analysis, data reduction, and visualization methods, we identify key features for accurate point and density forecasts. We evaluate various volatility models, including Generalized Autoregressive Conditional Heteroskedasticity (GARCH) and Stochastic Volatility (SV) models, to determine their forecasting performance. Our findings reveal significant heterogeneity in ETM volatility patterns, which challenge standard groupings by data providers and geological classifications. The results contribute to the literature on CRM economics and commodity volatility, offering novel insights into the complex dynamics of ETM markets and the modeling of their returns and volatilities.%
}%

\vspace{.5cm}

\small{\noindent\textbf{Key Words:} Critical Raw Materials; Energy Transition; Features; Volatility; Forecasting; Density forecasts.}

\newpage

\setcounter{footnote}{0}
\pagenumbering{arabic}
\doublespacing
\normalsize


\section{Introduction}\label{sec:intro}
\noindent The adoption of a cleaner energy mix, which is a cornerstone of policies to achieve net-zero greenhouse gas emissions by 2050, will require massive efforts to secure the metals needed to deploy technologies at scale. Renewable energy generation and the electrification of entire sectors of the global economy will drive demand for minerals such as copper, lithium, nickel, cobalt and rare earth elements \citep{worldbank2020minerals,iea2021netzero,iea2021role}. Metal prices are expected to reach historic highs for an unprecedentedly long period in a net-zero scenario \citep{boer2021energy}. While the total value of metals production would more than quadruple between 2021 and 2040 \citep{ECCRM2020}, current production and investment plans fall far short of what is needed to support the energy transition.

There is also widespread political recognition, at both national and supranational levels, of the growing strategic importance of minerals to economies around the world. Focusing on the EU, the European Commission (EC) regularly publishes a list of Critical Raw Materials (CRMs) and the political debate routinely addresses the issue of securing access to raw materials. CRMs are economically and strategically important materials characterized by a low degree of substitutability, high supply risk and production concentrated in a few countries \citep[see e.g.][]{graedel2015criticality}. These characteristics, coupled with low price elasticity of supply and demand, exacerbate the price volatility of these commodities \citep{boer2021energy,fally2018commodity}.

This paper represents the first attempt to model the price volatility of a broad set of Energy Transition Metals (ETMs), which are a subset of CRMs. ETMs have been included by the International Monetary Fund (IMF) in its Primary Commodity Price System with the aim of focusing on the price dynamics of metals critical to the energy transition. We first identify the essential data features for accurate point and density forecasts using a mix of exploratory data analysis, data reduction and visualisation methods \citep{kang2017visualising}. We then evaluate a number of volatility models, both in-sample and out-of-sample, and try to identify which data features are most important and which models best captures them. Our results show that ETMs have complex volatility dynamics, partly due to the illiquidity of the markets in which they are exchanged. 

Because some metals are often classified into homogeneous groups by data providers (e.g., base, precious, or minor metals), and some are mined together due to their natural occurrence in the same geological deposits (e.g., copper, nickel, and cobalt found in sulfide-rich ores), one might expect them to exhibit similar volatility patterns. On the contrary, when we focus on a key set of features and summarize them using data reduction techniques, we uncover significant heterogeneity. This analysis highlights that distinct clusters emerge, which neither align with the standard groupings used by data providers nor reflect the geological properties of metals. Moreover, we show that neither Generalized Autoregressive Conditional Heteroskedasticity (GARCH) models, which we use as a benchmark specification, nor a single Stochastic Volatility (SV) specification systematically prevails in a forecasting horse race.

Our paper intersects with two strands of literature. First, we contribute to the rapidly growing set of studies on the economics of CRMs and energy transitions \citep[see e.g.][]{boer2021energy,pommeret2022critical,diemer2022technology,graedel2015criticality,saadaoui2024ensuring}. This strand of literature remains highly fragmented, with different studies addressing very different aspects of CRMs and the transition to cleaner energy sources. Second, the paper also clearly belongs to the strand of literature that examines commodity volatility. While several papers have studied the volatility of commodities and other asset classes \citep[see e.g.][]{bakas2019volatility,bollerslev2018risk,chan2016modeling,ciner2020spillovers,dinh2022economic,han2022r,naeem2024tail,pham2024metals}, there is a notable gap regarding the volatility of ETMs.\footnote{An exception is \citet{BCGeneco}, who focus on the link between ETM returns and volatilities.} We contribute directly to this second branch of the literature.

Specifically, we evaluate the accuracy of different models in explaining and forecasting the returns and volatilities of a broad set of ETMs. In addition, we summarise the time series features of ETMs that can explain the heterogeneity in forecasting accuracy across volatility models. Feature extraction has mostly been applied in the statistical literature for clustering time series, selecting and combining forecasting models \citep{kang2017visualising,montero2020fforma,talagala2023meta}, but to our knowledge it is new in commodity forecasting.

The remainder of the paper is structured as follows: Section \ref{sec:datamet} describes the data and methodology. Section \ref{sec:results1} introduces the features space, while Section \ref{sec:results2} links features to the in-sample and out-of-sample evaluation of volatility models. Section \ref{sec:concl} concludes. Additional results and methodological details are presented in the Appendix.

\section{Data and Methods}\label{sec:datamet}
\subsection{Data and volatility proxies}\label{sec:data}
We focus on the volatility of the real prices of the 16 constituents of the ETM index published by the IMF.\footnote{The price index is then calculated as a weighted average of the prices of the 16 ETMs, where, according to the IMF, the weight assigned to each metal is based on its share of imports in the total imports of the 16 ETMs. The weights are taken from the ``Technical Documentation'' accompanying the IMF Primary Commodity Price Index (see: \url{https://www.imf.org/en/Research/commodity-prices}).} The ETM index consists of seven base metals (i.e., aluminum, cobalt, copper, lead, molybdenum, nickel, zinc), three precious metals (i.e., palladium, platinum, silver), and six minor metals (i.e., chromium, lithium, manganese, rare earth elements, silicon, vanadium) that are in high demand for technologies related to the energy transition. In the following, we refer to the minor metals as ``other ETMs''. The dataset in our analysis spans from July 2012 to December 2022 and consists of 126 monthly observations. See Table \ref{tab:Tab01Des} for further details.

The daily nominal prices of the ETMs are sourced from LSEG (formerly Refinitiv Eikon). Monthly real prices, $P_t$, are calculated by taking the average daily nominal prices deflated by the interpolated US Consumer Price Index sourced from BLS. More precisely, $P_t = D_t^{-1}\sum_{d=1}^{D_t} P_{d,t}$ where $P_{d,t}$ is the daily real price and $D_t$ represents the number of days in month $t$. This process also delivers daily real log-returns, $r_{d,t} = 100\times \log(P_{d,t}/P_{d-1,t})$, which we use to calculate various volatility proxies as described below.

There are two reasons for using real prices at monthly sampling frequency rather than considering weekly or daily data. Firstly, the low liquidity in certain minerals markets makes the use of daily data impractical. Secondly, since true volatility is not directly observable, we require a volatility proxy to assess the predictive ability of models. In this paper, we construct such proxies by aggregating daily data to a monthly frequency \citep{schwert1989}.

\subsection{Volatility models}\label{sec:models}
We compare two classes of volatility models: GARCH and SV. A key difference between GARCH and SV models is that the latter attach a stochastic error component to the volatility equation. While this makes SV models capable of capturing complex volatility dynamics, it also renders estimation more involved. We rely on Bayesian methods for estimating SV models.\footnote{We estimate GARCH models with maximum likelihood. The Bayesian estimation of SV models and the computation of the marginal likelihood are carried out using the code in \citet{chan2016modeling}. For the sake of brevity, we present further details and assumptions for the estimation of SV models in a Supplement available upon request from the authors.}

The GARCH(1,1) model of \citet[]{bollerslev1986generalized} can be written as follows:
\begin{align}
r_t & = \mu + \epsilon_t, \quad \epsilon_t \sim N(0, \sigma_t^2) \\
\sigma_t^2 & = \alpha_0 + \alpha_1 \epsilon_{t-1}^2 + \beta_1 \sigma_{t-1}^2 \label{eq:garch}
\end{align}
where innovations $\epsilon_t$ are normally distributed with zero mean and conditional variance $\sigma_t^2$. The parameters in the conditional variance equation should satisfy $\alpha_0>0$, $\alpha_1>0$, $\beta_1\geq 0$ to guarantee that $\sigma_t^2\geq 0$. Moreover, the model is covariance-stationary if $\alpha_1 + \beta_1<1$. GARCH(1,1) forecasts serve as a benchmark against which we compare five different SV specifications aimed at capturing different features of returns and volatilities of the ETMs.

The simplest SV model we analyze, denoted as SV(1), assumes that the log-volatility $\log{\sigma_t^2}$ evolves as a stationary autoregressive process of order 1, $AR(1)$:
\begin{align}\label{eq:volatility}
r_t & = \mu +\epsilon_t, \qquad \epsilon_t\sim N(0, e^{\log{\sigma_t^2}})\\
\log{\sigma_t^2} & = \phi_0 + \phi_1(\log{\sigma_{t-1}^2}-\phi_0)+\eta_t, \qquad \eta_t \sim N(0,\omega^2)
\end{align}
with $|\phi_1|<1$ to ensure covariance-stationarity of the AR(1) process. The volatility equation incorporates a stochastic error term $\eta_t$, which is normally distributed with zero mean and a variance of $\omega^2$.

A simple way to extend the SV(1) model is to allow the log-volatility to follow an AR(2), so that we can capture richer variance dynamics:
\begin{equation}
\log{\sigma_t^2} = \phi_0 + \phi_1(\log{\sigma_{t-1}^2}-\phi_0)+ \phi_2(\log{\sigma_{t-2}^2}-\phi_0)+\eta_t, \quad \eta_t \sim N(0,\omega^2)
\end{equation}
in the SV(2) model, we assume that the roots of the characteristic polynomial associated with $\left(\phi_1,\phi_2\right)$ lie outside the unit circle.

We also examine is the SV model with $t$-distributed innovations, SV(1)-t, which allows for more extreme observations compared to SV(1). In the SV(1)-$t$, the log-volatility dynamics are again approximated by an AR(1) process, but now we assume that innovations in the return distribution are $t$-distributed with $\nu$ degrees of freedom: $\epsilon_t \sim t_\nu(0,e^{\log\sigma^2_t})$.

Next we consider the SV model with leverage, SV(1)-L, that allows for asymmetric effects of positive and negative returns on volatility. Log-volatility is approximated by an AR(1) process as in the SV(1) model, but now we assume that returns innovations, $\epsilon_t$, and volatility innovations, $\nu_t$, follow a bivariate normal distribution:
\begin{equation}
\begin{bmatrix}
\epsilon_t \\
\eta_t
\end{bmatrix} \sim N%
\left(%
\begin{bmatrix}
0\\0
\end{bmatrix},%
\begin{bmatrix}
\sigma_t^2 & \rho \sigma_t \omega \\
\rho \sigma_t \omega & \omega^2
\end{bmatrix} \right)
\end{equation}
where $\rho$ a correlation coefficient. When $\rho<0$ and a negative shock hits returns at time $t$, volatility at time $t + 1$ will increase; when $\rho=0$ the SV(1)-L boils down to the standard SV(1) model.

Lastly, we consider a model with jumps, denoted as SV(1)-J, to capture infrequent, but significant price movements. In this case, log-volatility continues to be approximated by an AR(1) process, but the equation for returns becomes:
\begin{equation}
r_t = \mu + k_t q_t+\epsilon_t
\end{equation}
where $q_t \in \{0,1\}$ is a jump variable with success probability $\mathbb{P}(q_t=1)=\kappa$ and $k_t\sim N(\mu_k,\sigma_k^2)$ determines the size of the jump if it occurs at time $t$. As for the innovations, we continue assuming that: $\epsilon_t \sim N(0, e^{\log{\sigma_t^2}})$.

\subsection{The forecasting exercise}\label{sec:foreex}
We forecast volatilities and returns one step ahead using an expanding window approach. We first estimate models using data from July 2012 to December 2017 and forecast returns and volatilities for January 2018. We then add one observation to the estimation sample and forecast returns and volatilities for February 2018. This process is repeated until the end of the sample, resulting in a total of 60 one-step-ahead forecasts from January 2018 to December 2022.\footnote{A rolling window approach, where forecasts are generated using a sample of 66 observations that is moved forward one period each time we generate a new forecast, yields qualitatively similar results.}

We evaluate models for their ability to produce accurate one-step-ahead point forecasts of volatilities, and density forecasts of returns. For the GARCH(1,1) model, point forecasts of the conditional variance, $\hat{\sigma}^2_{t+1|t}$, are obtained by iterating Equation \ref{eq:garch} forward by one period. In the case of SV models, the one-step-ahead point forecast $\hat{\sigma}^2_{t+1|t}$, is the mean of the predictive distribution of the conditional variance. As for the density forecasts of returns, SV models are estimated using Bayesian methods and therefore readily provide the predictive distribution of both returns and volatilities. For the GARCH(1,1) model, density forecasts of returns are obtained by drawing from the distribution of error terms given the volatility forecasts and ignoring parameter uncertainty \citep{Bassetti2020}.

\bigskip

\noindent\textit{Volatility proxies}. While the true volatility of ETM prices is not directly observable, to evaluate forecasts we need an unbiased estimator of the latent variable of interest, namely conditional volatility \citep[see e.g.][]{hansen2005forecast,patton2011volatility}. We rely on two of such proxies: realized volatility (RV) and the adjusted squared range ($RG2^*$). Annualized monthly RV is defined as:
\begin{equation}
    RV_t=252\times\frac{1}{D_t}\sum_{d=1}^{D_t} r_{d,t}^2
\end{equation}
The annualized adjusted squared range is frequently employed as an alternative volatility proxy and can be written as \citep[see e.g.][]{garman1980estimation,patton2011volatility}:
\begin{equation}
    RG2^*_t=1200\times\left(\frac{RG_t}{2\sqrt{\log(2)}}\right)^2
\end{equation}
where $RG_t = \underset{d\in t}{\max} (p_{d,t}) - \underset{d\in t}{\min}( p_{d,t})
$, represents the difference between the maximum and minimum real log-price recorded in month $t$.

\bigskip

\begin{figure}
    \centering
    \caption{Mean Squared Forecast Error and QLIKE loss}
    \includegraphics[width=\textwidth]{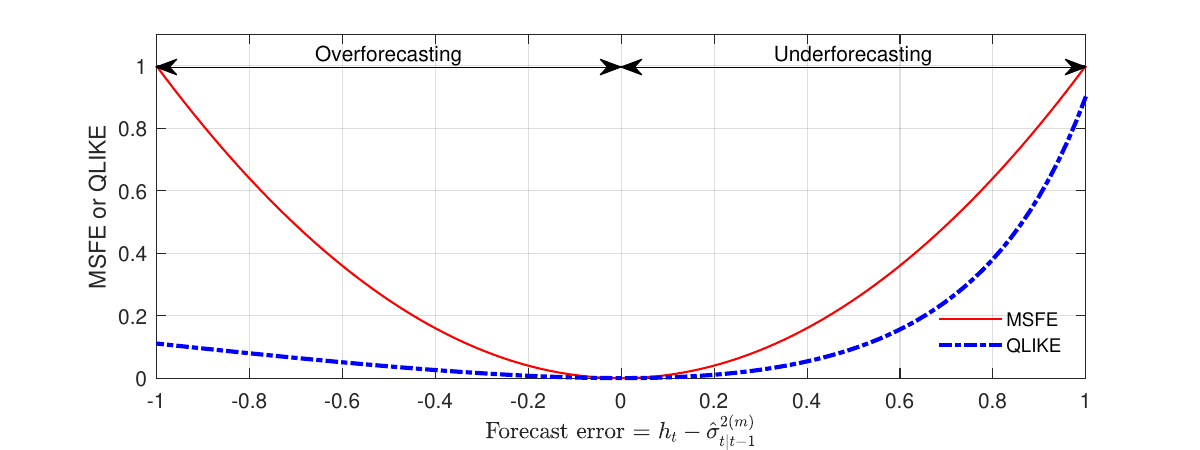}
    \label{fig:FigQlikeMSE}
\end{figure}

\noindent\textit{Loss functions.} We evaluate point forecasts of the conditional variance with two alternative loss functions, the Root Mean Squared Forecast Error (RMSFE) and the QLIKE. These are defined as follows:
\begin{align}
    RMSFE^{(m)} & = \sqrt{\frac{1}{P}\sum_{t=1}^P\left(h_t-\hat{\sigma}_{t|t-1}^{2(m)}\right)^2}\\
   QLIKE^{(m)} & = \frac{1}{P}\sum_{t=1}^P\left[\frac{h_t}{\hat{\sigma}_{t|t-1}^{2(m)}}-\log\left(\frac{h_t}{\hat{\sigma}_{t|t-1}^{2(m)}}\right)-1\right]
\end{align}
where $P$ denotes the size of the evaluation sample, $h_t$ represents one of the volatility proxies (i.e. $RV_t$ or $RG2^*$), and $\hat{\sigma}_{t|t-1}^{2(m)}$ is the one-step-ahead volatility forecast from model $m$. As shown by \cite{patton2011volatility}, both the MSFE and the QLIKE are robust loss functions, which implies that the ranking of competing volatility forecasts is not affected by the use of alternative volatility proxies. As we can see from Figure \ref{fig:FigQlikeMSE}, contrary to the MSFE, which is a symmetric loss function, the QLIKE criterion penalizes underforecasting more heavily than overforecasting.

\FloatBarrier

Moving on to density forecasts of returns, we consider the quantile-based continuous ranked probability score (qCRPS) of \citet{gneiting2011comparing}. The interpretation of scoring rules is the same as that given to loss functions: in a pairwise comparison, the model with the lowest score is ranked as the most accurate. The qCRPS is defined as (we drop the model superscript, $m$, for ease of notation): 
\begin{equation}
    \widehat{QS}_{t}=\frac{1}{J-1} \sum_{j=1}^{J-1} \widehat{QS}_{t}^{\alpha_j}= \frac{1}{J-1} \sum_{j=1}^{J-1} 2\left[ \mathbb{I}\left(r_{t+1}\leq \hat{q}^{\alpha_j}_{t+h|t}\right) - \alpha_j \right]\times\left(\hat{q}^{\alpha_j}_{t+h|t}-r_{t+1}\right),
    \label{eq:crps}
\end{equation}
where $\mathbb{I}(\cdot)$ denotes the indicator function and $\hat{q}^{\alpha_j}_{t+h|t}$ is the $h$-step ahead quantile forecast for $r_{t+1}$ at level $\alpha_j=j/J$ with $J=20$, which corresponds to $\alpha_j = 0.05, 0.10, \ldots, 0.95$. We focus on weighted versions of the qCRPS:
\begin{align}
    \widehat{wQS}_t & = \frac{1}{J-1} \sum_{j=1}^{J-1} \pi(\alpha_j) \widehat{QS}_{t}^{\alpha_j}
\end{align}
where we select weights, $\pi(\alpha_j)$, to emphasize the center $\pi(\alpha_j)=\alpha_j(1-\alpha_j)$ or the tails of the distribution $\pi(\alpha_j)=(2\alpha_j-1)^2$.

\section{The features of ETM time series}\label{sec:results1}
\subsection{Descriptive statistics}
\begin{figure}[t]
    \centering
    \caption{Evolution of prices and volatilities over time}
    \includegraphics[width=\textwidth]{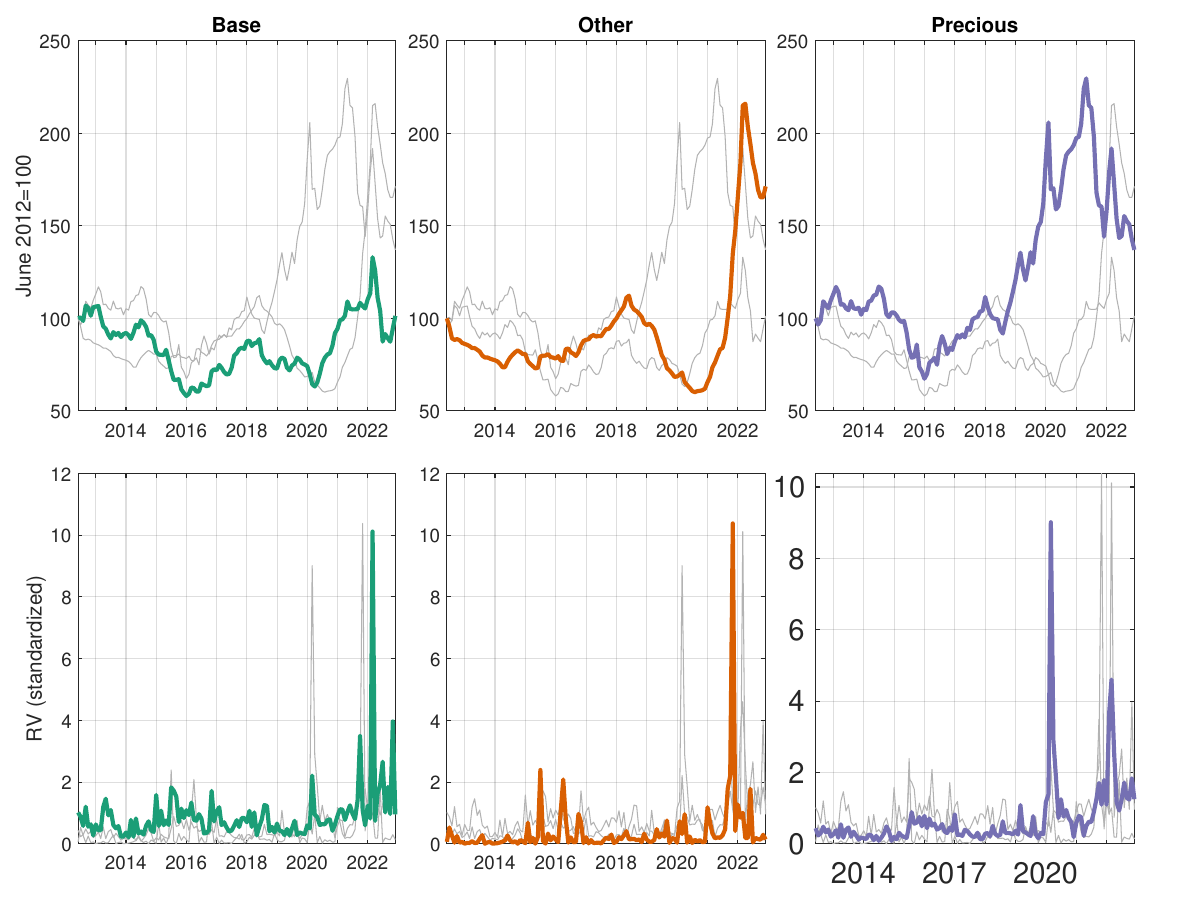}
    \label{fig:Fig02Desc}
\end{figure}

As can be seen from Table \ref{tab:Tab01Des}, we group ETMs into three broad categories: ``base'', ``other'' and ``precious''. Green energy technologies, hydrogen and electric mobility all use many ETMs. The projected increases in demand in 2050, as implied by the IEA's net-zero emissions scenario, are in some cases of huge order of magnitude compared to 2020 levels. The last column shows the increase in price volatility, as measured by the coefficient of variation. This highlights that the period 2016-22 was a period of very high price volatility, particularly for base and precious metals. Figure \ref{fig:Fig02Desc} shows the evolution of prices and RV for different categories of metals, while plots of the series for individual metals appear in Section \ref{sec:appfigtab} of the Appendix.

\newcolumntype{C}[1]{>{\scriptsize\hsize=#1\hsize\centering\arraybackslash}X}
\newcolumntype{L}[1]{>{\scriptsize\hsize=#1\hsize\raggedright\arraybackslash}X}

\begin{table}[t]
\caption{Technologies, expected demand increases and volatility of ETMs}
\begin{tabularx}{\textwidth}{L{1.2}|C{.9}|C{.8}C{.8}C{.6}C{.8}C{.8}C{.8}|C{1.3}|C{1}}
\hline
 & & \multicolumn{6}{c|}{\scriptsize Tecnology} & Projected & \\\cline{3-8}
Metal      & Category      & Solar        & Wind         & EV           & Battery      & Hydrogen     & Other        &  Demand$^{\frac{2050}{2022}}$ & cv$^{\frac{2016-22}{2012-15}}$ \\\hline
Aluminum  & Base     &              &              &              &              &              & $\checkmark$ & 2.0                           & 1.82                        \\
Cobalt     & Base     &              &              & $\checkmark$ & $\checkmark$ & $\checkmark$ & $\checkmark$ & 4.3                           & 4.76                        \\
Copper     & Base     & $\checkmark$ & $\checkmark$ & $\checkmark$ & $\checkmark$ & $\checkmark$ & $\checkmark$ & 3.0                           & 1.30                        \\
Lead       & Base     & $\checkmark$ &              &              &              &              & $\checkmark$ & 1.3                           & 1.28                        \\
Molybdenum & Base     & $\checkmark$ & $\checkmark$ &              &              &              & $\checkmark$ & 2.5                           & 1.48                        \\
Nickel     & Base     & $\checkmark$ & $\checkmark$ & $\checkmark$ & $\checkmark$ & $\checkmark$ & $\checkmark$ & 8.2                           & 1.72                        \\
Zinc       & Base     & $\checkmark$ & $\checkmark$ &              &              &              & $\checkmark$ & 3.0                           & 1.82                        \\\hline
Chromium   & Other    &              & $\checkmark$ &              &              &              & $\checkmark$ & 2.5                           & 1.59                        \\
Lithium    & Other    &              &              &              & $\checkmark$ &              & $\checkmark$ & 16.1                          & 20.13                       \\
Manganese  & Other    &              & $\checkmark$ & $\checkmark$ & $\checkmark$ &              & $\checkmark$ & 12.9                          & 3.83                        \\
REE        & Other    &              & $\checkmark$ & $\checkmark$ &              & $\checkmark$ &              & 175.4                         & 1.43                        \\
Silicon    & Other    & $\checkmark$ &              & $\checkmark$ & $\checkmark$ &              &              & 0.1                           & 5.13                        \\
Vanadium   & Other    &              &              &              & $\checkmark$ &              &              & -                             & 3.45                        \\\hline
PGMs       & Precious &              &              &              &              & $\checkmark$ &              & 119.5                         & 0.65                        \\
Silver     & Precious & $\checkmark$ &              &              &              &              & $\checkmark$ & 0.4                           & 3.83                       \\\hline
\end{tabularx}
\label{tab:Tab01Des}
\caption*{\scriptsize \textit{Notes}: in column 1, palladium and platinum are included in the Platinum Group Metals (PGMs). The technologies in columns 3-8 and the projected increase in demand (i.e. projected demand in 2050 divided by demand in 2022) in column 9 are taken from \cite{ieacrm} and are based on the ``Net Zero Emission Scenario by 2050''. For vanadium we do not show the projected increase in demand as there is no demand in 2022. The last column show the ratio of the coefficient of variation of prices in 2012-15 to that in 2016-22. For PGMs, the average CV ratio for palladium and platinum is shown. The category 'other' includes low-emission power generation technologies (e.g. hydropower, geothermal, bioenergy for electricity and nuclear power) and electricity networks.}
\end{table}  

\subsection{A principal component analysis of features}\label{sec:feat}
To illustrate some stylized facts about the volatility of ETM prices, we focus on a few ``features'' of RV time series. While a ``feature'' refers to any measurable characteristic of a time series, their precise definition and selection is problem-specific \citep{talagala2023meta}. In our analysis, we examine features of realized volatility classified into three primary groups: autocorrelation (ACF) features, distributional features, and other features.\footnote{We use functions implemented in the \texttt{tsfeatures} R package by \citet{hyndmanR} to compute some of these features.}

In the group of ACF features, we consider ($F_1$) the first-order autocorrelation coefficient of RV $\widehat{ACF}(1)$; ($F_2$) the sum of squares of the first ten autocorrelation coefficients of RV, $\sum_{j=1}^{10}\widehat{ACF}(j)^2$; ($F_3$) the Hurst coefficient, which serves as a measure of ``long memory'' and thus, broadly speaking, signals the presence of significant autocorrelations over many lags.\footnote{The Hurst coefficient or exponent $H$ is estimated by fitting the equation $E[\frac{R(n)}{S(n)}]=Cn^H$ to the data, where $R$ represents the range of cumulative deviates of the data, $S$ denotes the standard deviation of the data, $n$ is the number of observations, and $C$ is a constant. The estimation process involves taking the logarithm of both sides of the equation and subsequently fitting a straight line to the transformed data. The slope of this line corresponds to the value of $H$. The Hurst coefficient indicates the behavior of the trend in a time series, whether it is persistent (Hurst coefficient $> 1/2$), antipersistent (Hurst coefficient $< 1/2$), or exhibits short memory (Hurst coefficient $= 1/2$). In the case of persistence, a high value is more likely to be followed by another high value. Conversely, in the case of antipersistence, a single high value is expected to be followed by a low value, with the subsequent value likely to be high.} As far as distributional characteristics are concerned, we analyze the sample skewness ($F_4$) and kurtosis ($F_5$) of the RV, as well as the percentage of zero returns as a proxy for illiquidity ($F_6$). Finally, in the group of other features, we analyse the Shannon spectral entropy measure ($F_7$) and a proxy for the ``spikiness'' ($F_8$) of RV \citep[see, e.g.,][]{rob2018forecasting}. The Shannon spectral entropy captures the noise level and the forecastability of the RV. A value close to 1 indicates high noise and makes forecasting difficult, while a value close to 0 indicates a strong trend (or seasonality), which makes forecasting easier. On the other hand, the ``spikiness'' of the RV measures the prevalence of spikes in the remainder component of the STL decomposition of the series.\footnote{The STL (Seasonal and Trend decomposition using Loess) decomposition method decomposes a series, denoted as $y_t$, into three distinct components: the trend component, represented as $T_t$; the seasonal component, denoted as $S_t$; and the remainder component, represented as $R_t$. Mathematically, the decomposition can be expressed as follows: $y_t=T_t+S_t+R_t$. The spikiness is computed as the variance of the leave-one-out variances of $R_t$.} 

In general, the stronger the ACF features, the easier it should be for models to use historical patterns to accurately describe the stylized facts of ETM price volatility. On the other hand, the group of distributional features (i.e. skewness, kurtosis and percentage of zeros) and the group of other features (i.e. spectral entropy and spikiness) indicate the presence of outliers and noise in the time series. Consequently, a strong presence of these features indicates that it is more difficult for models to exploit past patterns and accurately model volatility.

\begin{figure}[ht]
    \centering
     \caption{Features of Realized Volatility}
     \includegraphics[width=0.99\textwidth]{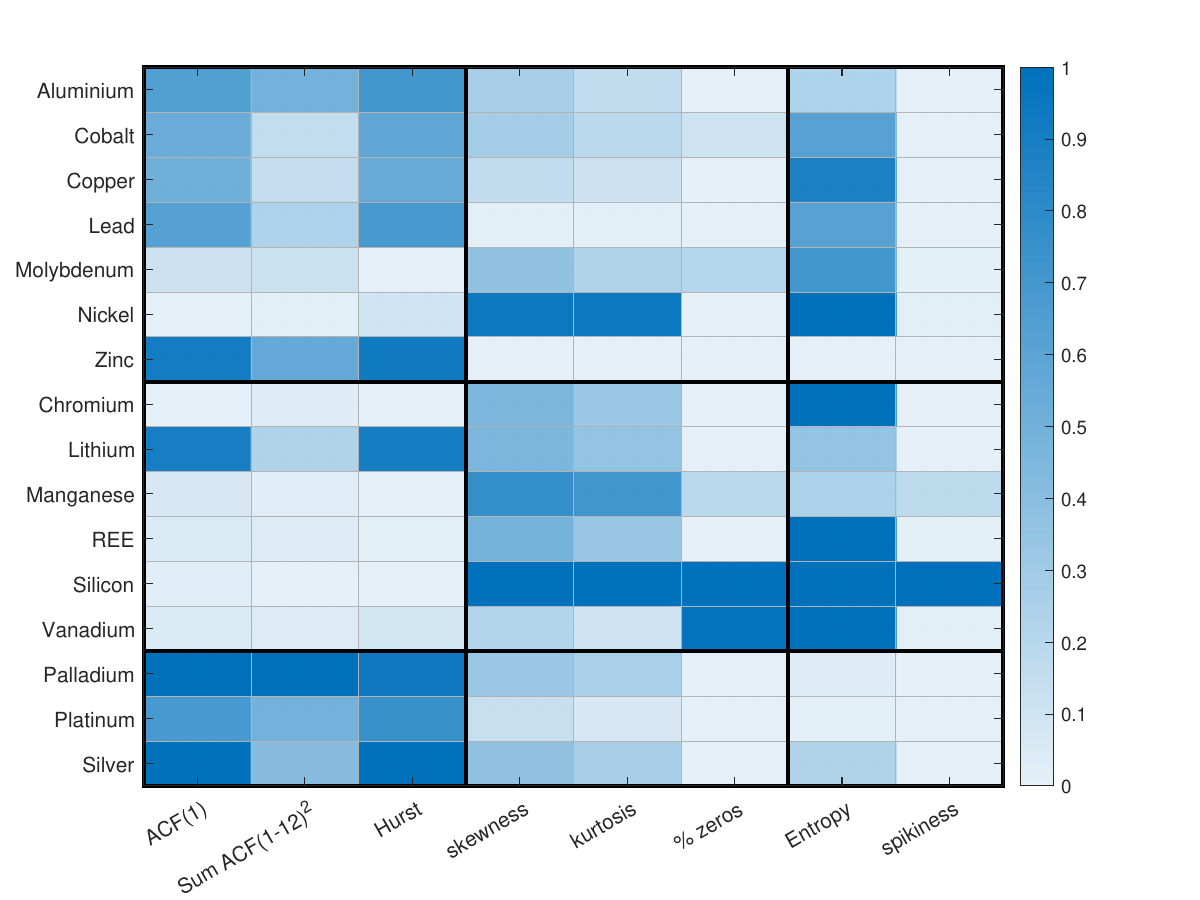}
    \label{fig:featheatmap}
         \caption*{\scriptsize\textit{Notes}: the heatmap shows the features of the RVs of the different metals. The thick black lines divide the metals into base, other and precious and the features into ACF, distributional and other. All feaures are normalised in the range (0,1). Darker shades indicate higher values of the feature, as shown in the colour bar.}
\end{figure}

After computing the eight features, we normalize them to the $(0,1)$ interval\footnote{For ACF(1), we first take the absolute value.} and then collect them in the feature vector $\mathbf{F} = \left[F_1, F_2, \ldots, F_8\right]^{\prime}$. In Figure \ref{fig:featheatmap}, we show a heatmap with the normalized feature values for different ETMs and the eight features. The bold black lines separate distinct clusters of ETMs (base, precious, other) and features (i.e., ACF, distributional, other). Base and precious metals tend to have high values for ACF features and low values for spikiness and zero returns. On the contrary, most metals in the ``other'' category are characterized by low values for ACF features and high values concentrated on distributional features and entropy. Silicon is a case in point with high values for skewness, kurtosis, spectral entropy, spikiness and a high percentage of zero returns.

\bigskip

\subsection{Representing ETMs in the feature space}\label{sec:featspace}
Following \citet{kang2017visualising}, we use Principal Component Analysis (PCA) to represent each time series of RV as a point in a two-dimensional ``feature space'', also referred to as instance space. Together, the first two principal components explain 81.3\% of the data variance, with the first component, $PC_1$, accounting for 66.2\% of the total. As we can see from Equation \eqref{eq:pca} $PC_1$ increases with the ACF features and decrease with the other features.

\begin{figure}[ht]
    \centering
     \caption{$R^2$ between principal components and features and the feature space of RVs.}
     \includegraphics[width=0.495\textwidth]{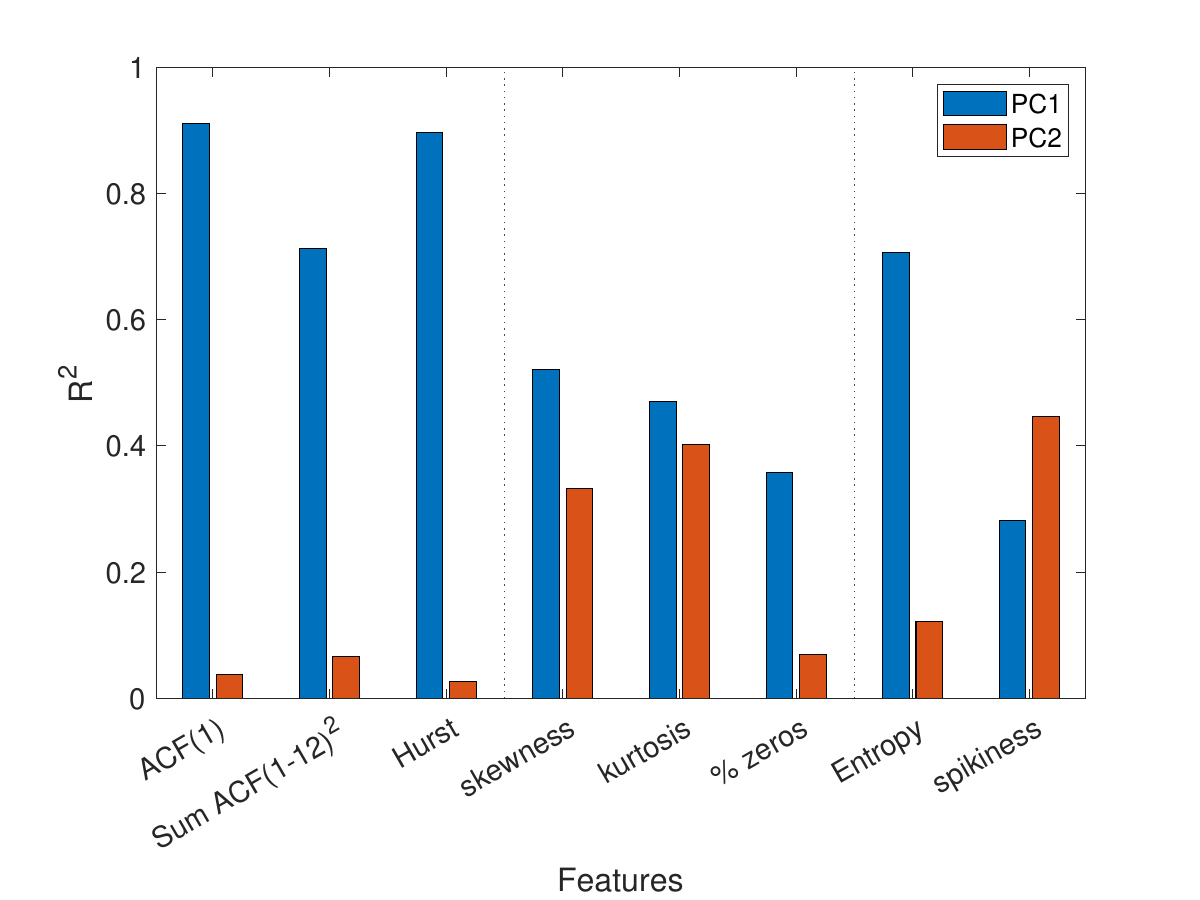}%
     \includegraphics[width=0.495\textwidth]{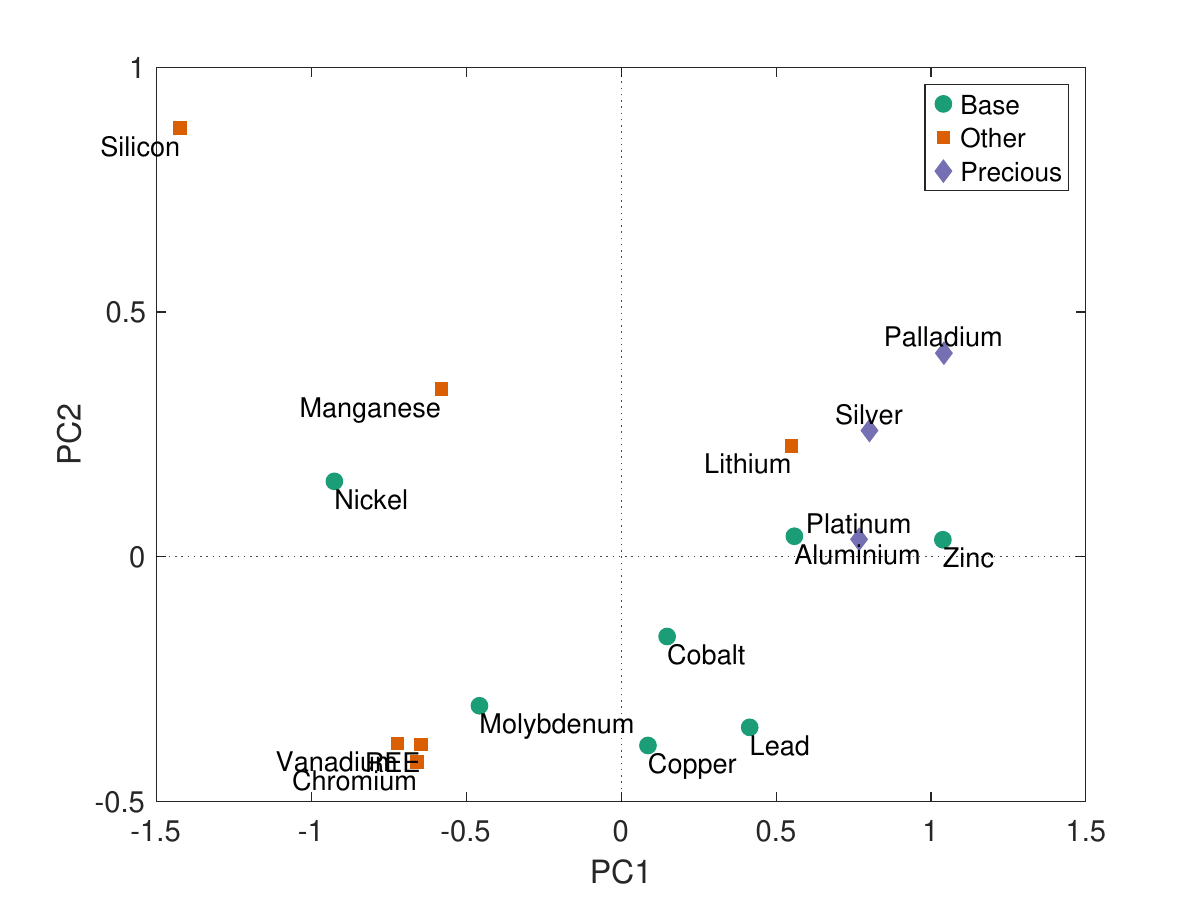}%
    \label{fig:features}
     \caption*{\scriptsize\textit{Notes}: the left-hand panel shows the $R^2$ values from regressions of the 8 individual RV time series separately against $PC_1$ and $PC_2$. The righ-hand panel shows the instance space of RV time series. $PC_1$ and $PC_2$ are the first two principal components, projected from the eight-dimensional feature space.}
\end{figure}

\begin{equation}
\begin{bmatrix}
    PC_1\\ 
    PC_2
\end{bmatrix} =
\begin{bmatrix}
\begin{array}{rrr:rrr:rr}
 0.48 &  0.30 &  0.50 & -0.28 & -0.27 & -0.26 & -0.43 & -0.17 \\
 0.21 &  0.19 &  0.18 &  0.46 &  0.53 &  0.24 & -0.37 &  0.45
\end{array}
\end{bmatrix}
\mathbf{F}
\label{eq:pca}
\end{equation}

\noindent The left panel of Figure \ref{fig:features} illustrates the $R^2$ values from regressions of the 8 individual RV time series separately against $PC_1$ and $PC_2$. The $R^2$ values, represented as bar charts, show that $PC_1$ loads primarily on ACF features, while $PC_2$ on distributional features, spectral entropy and spikiness.

The scatter plot in the right-hand panel of Figure \ref{fig:features} provides further interesting information. Focusing on the horizontal axis, we can see that  $PC_1$ clearly separates the series into two clusters, the first consisting mainly of precious and base metals and the second mainly of other metals.  $PC_2$ further separates the series and shows that the RV of silicon has distinctive features compared to other metals.

To better understand the relationship between the characteristics of the series and their division into clusters, Figure 3 shows both the positioning of the ETMs and the distribution of the characteristic values in the instance space. The size of the bubbles in the graphs is proportional to the value of the features.

The first three graphs show that all series associated with positive values of the first principal component, namely those in the right part of the instance space, have high values of the ACF features.

\begin{figure}[ht]
    \centering
     \caption{Instance space and feature values}
     \includegraphics[width=0.45\textwidth]{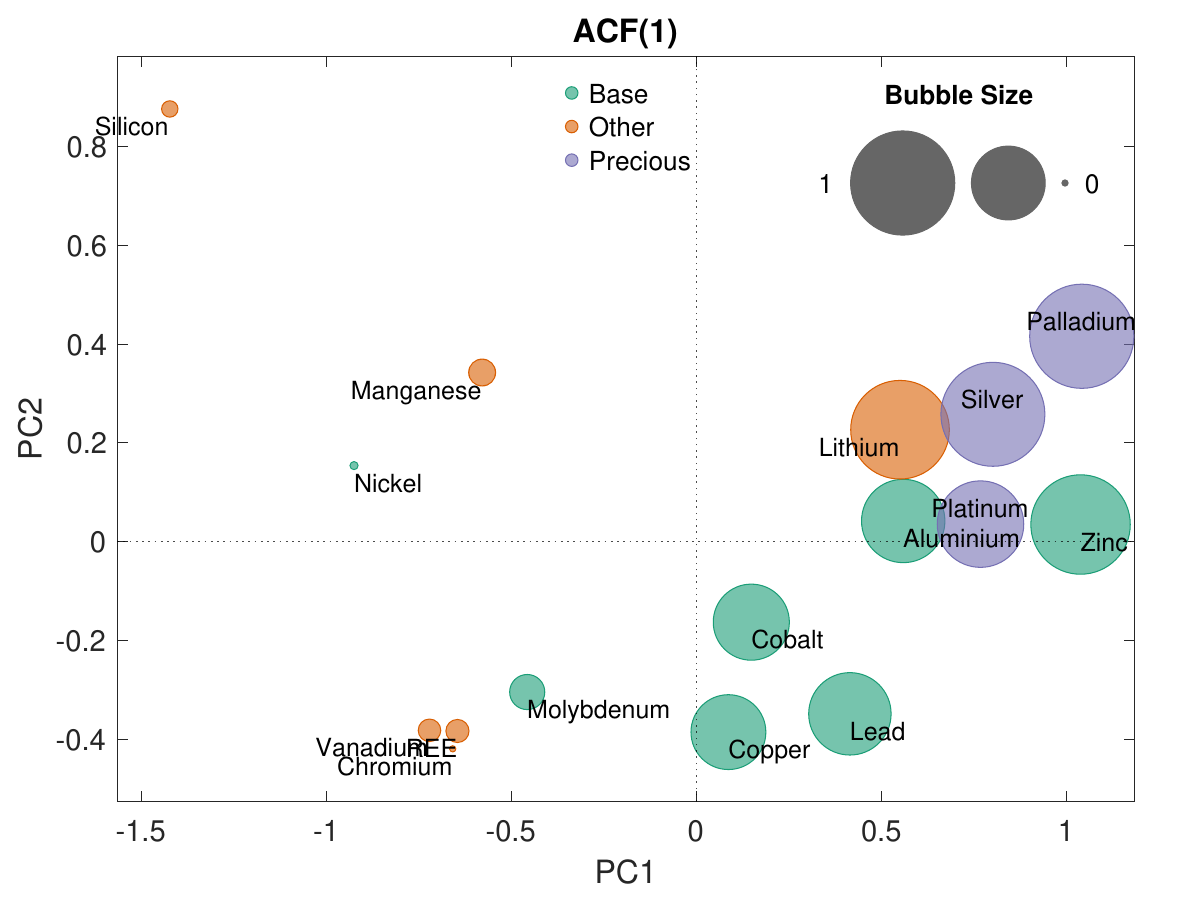}%
     \includegraphics[width=0.45\textwidth]{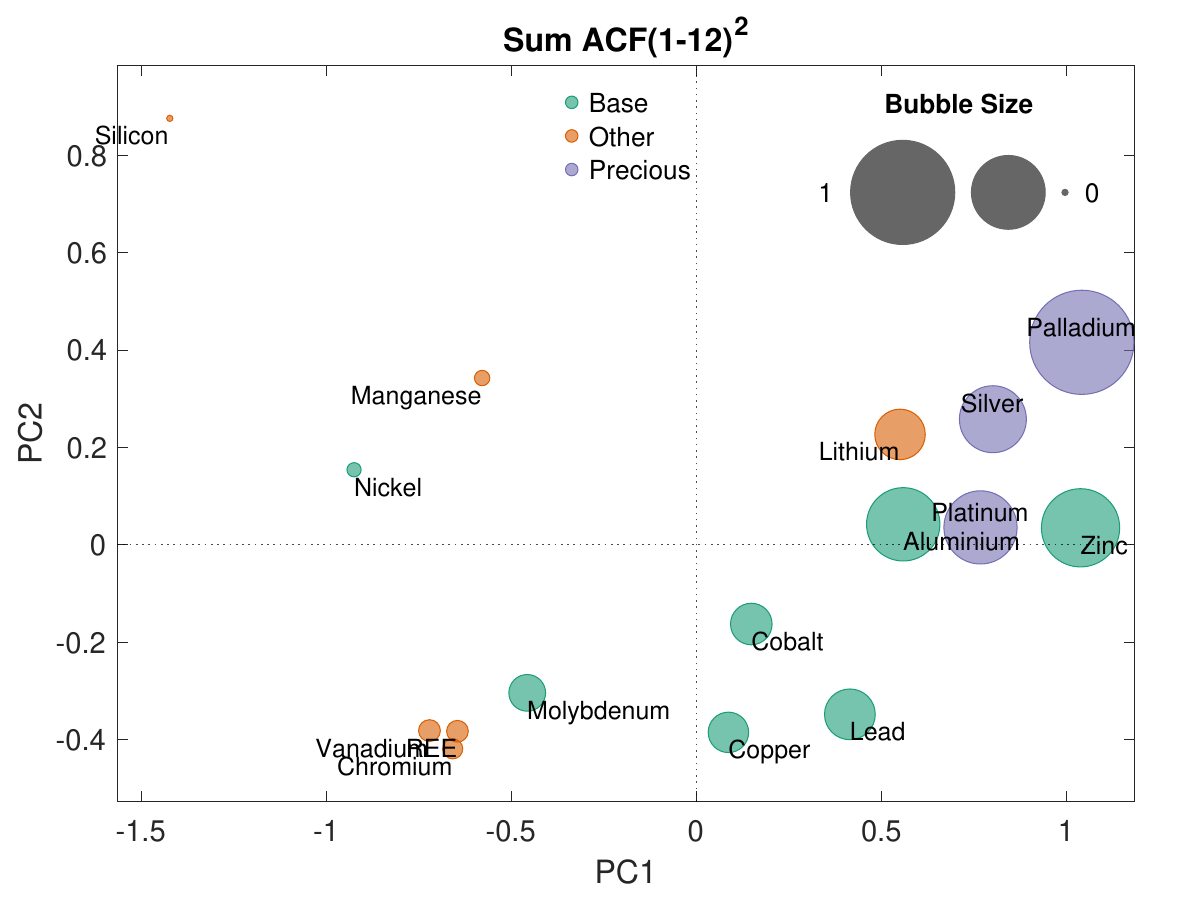}\\
     \includegraphics[width=0.45\textwidth]{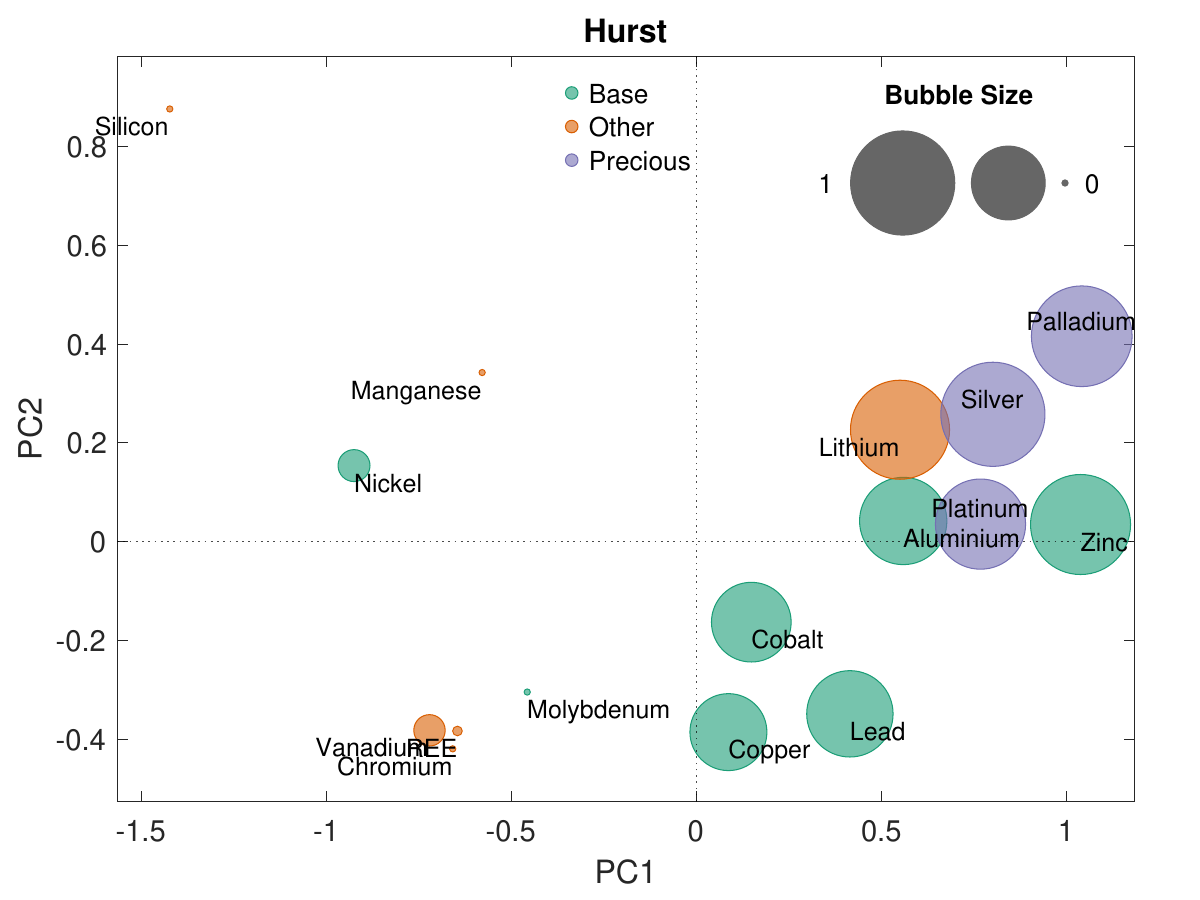}%
     \includegraphics[width=0.45\textwidth]{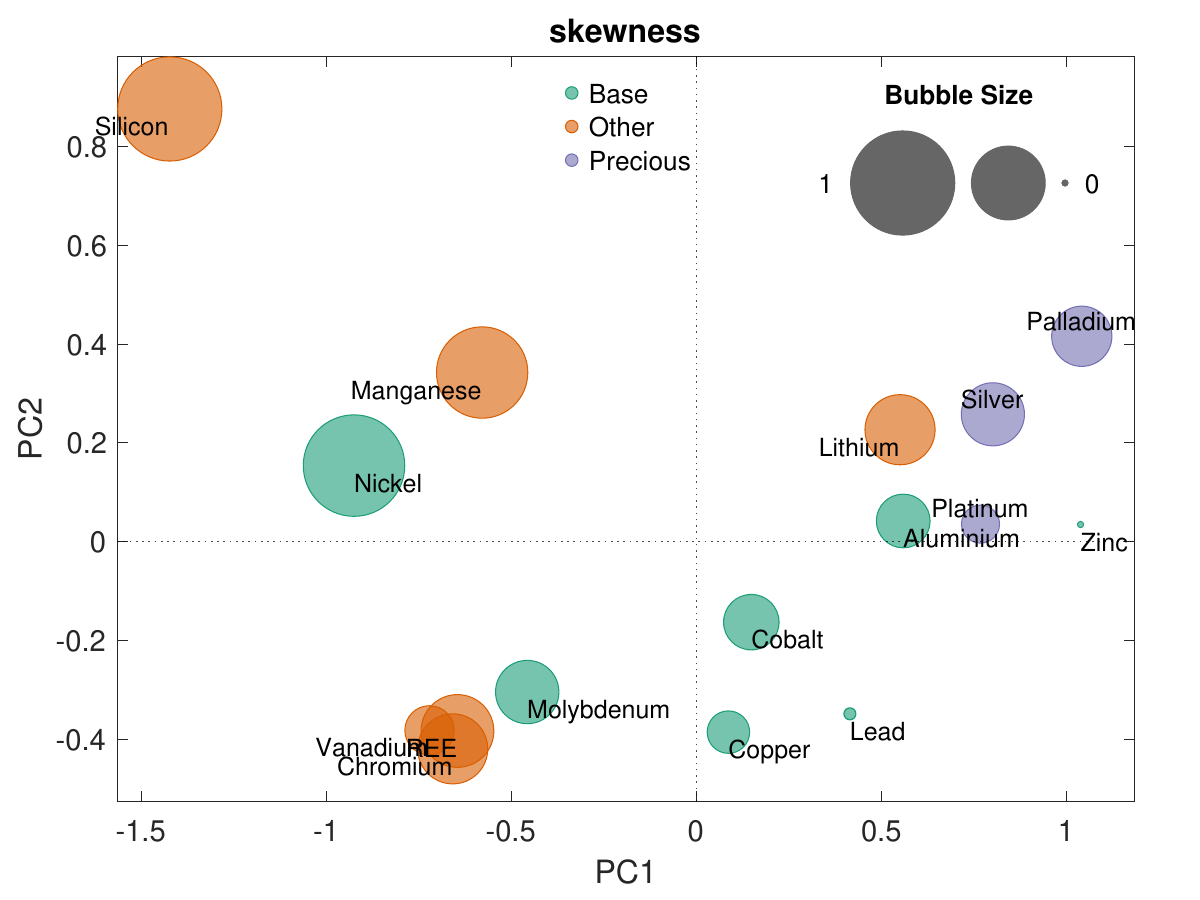}\\
     \includegraphics[width=0.45\textwidth]{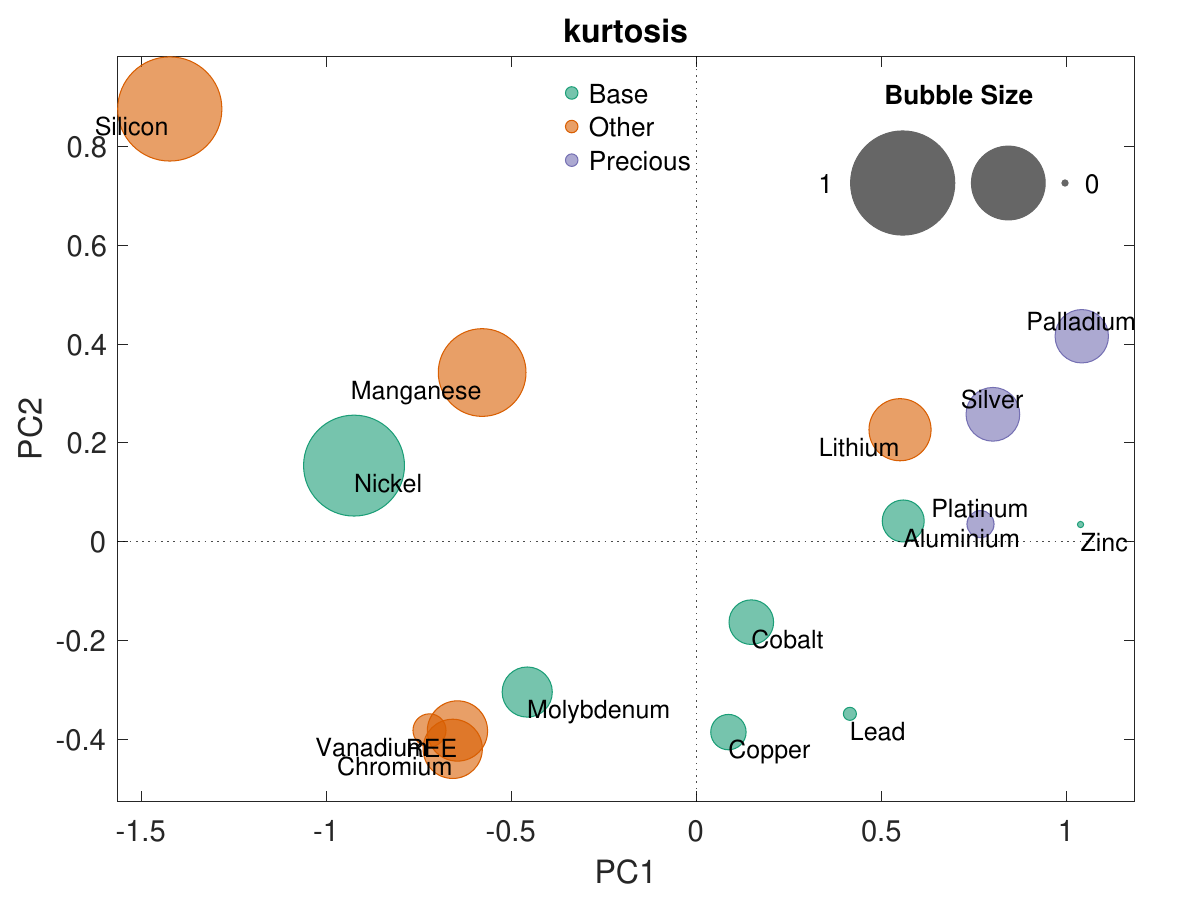}%
     \includegraphics[width=0.45\textwidth]{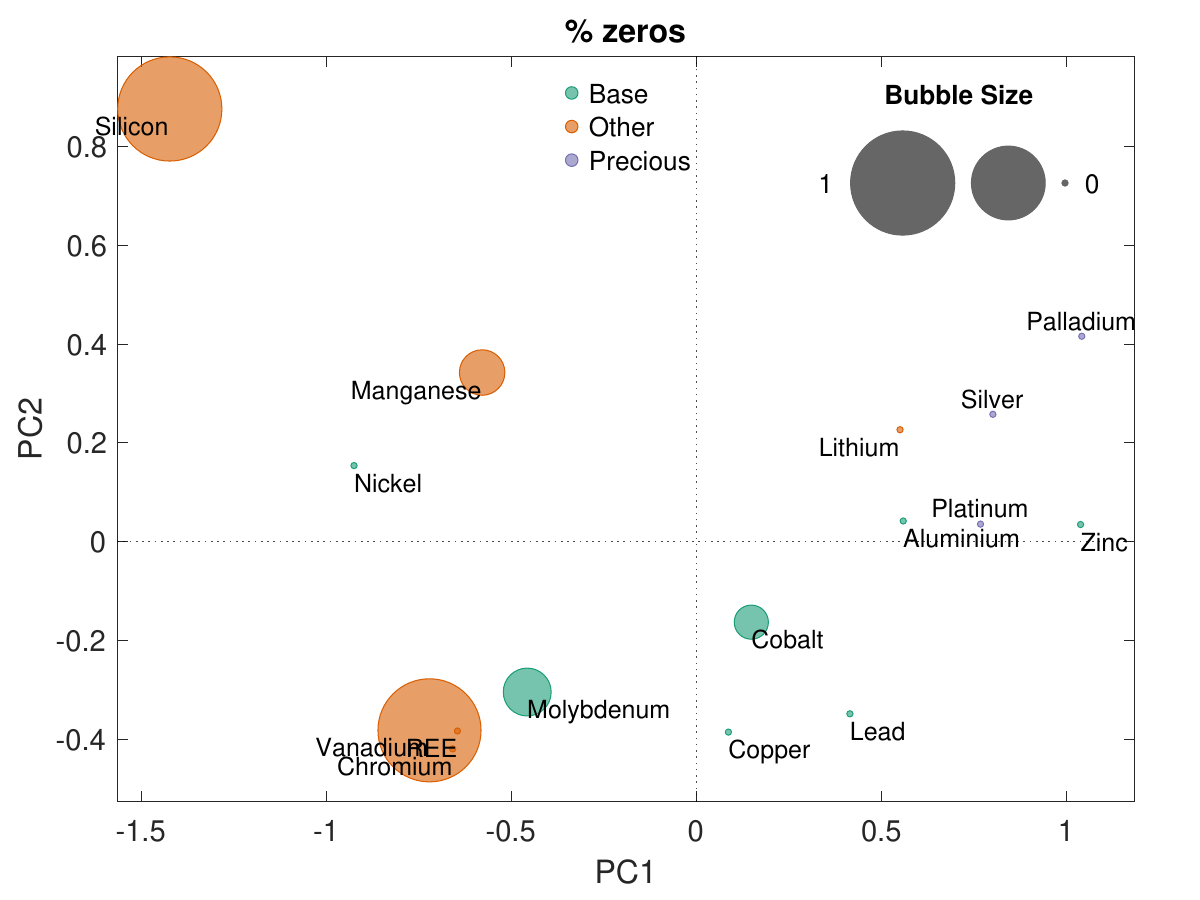}\\
     \includegraphics[width=0.45\textwidth]{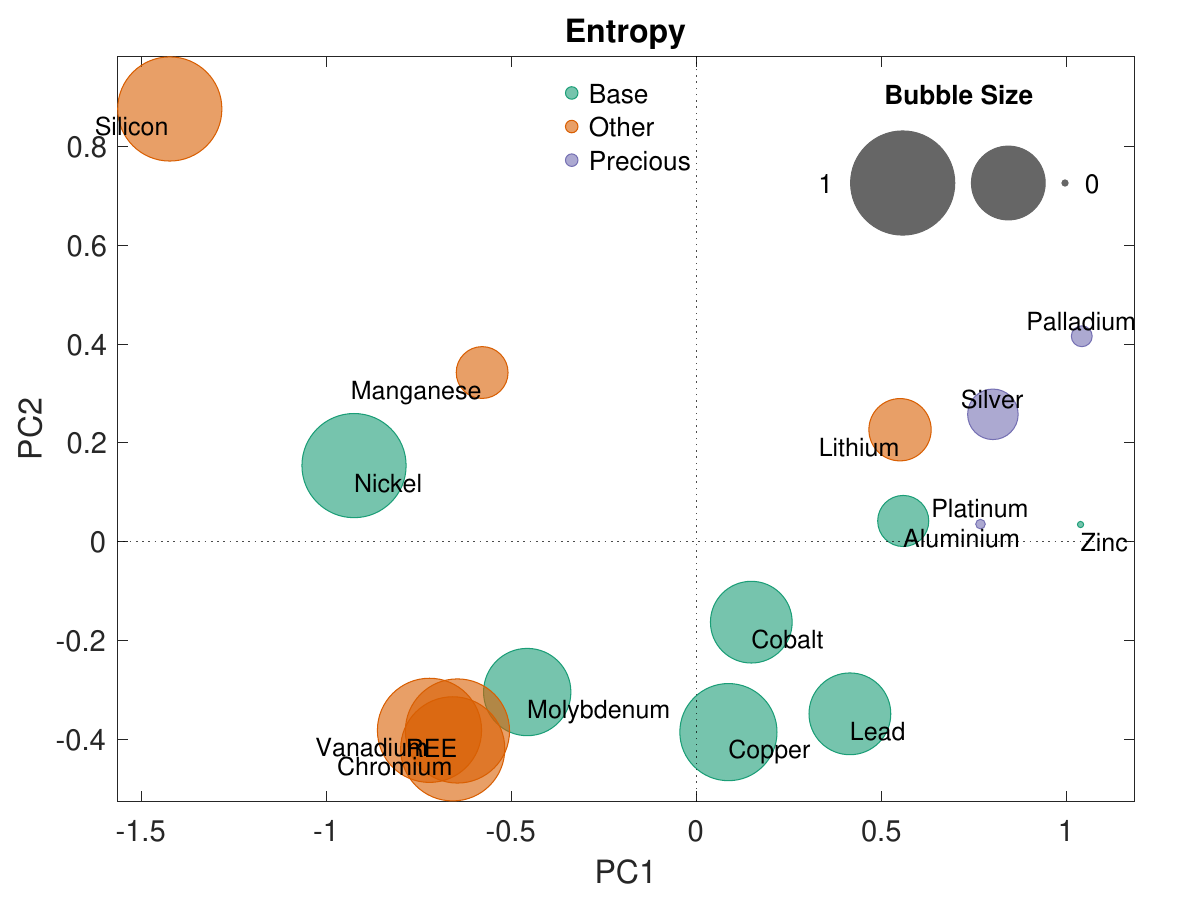}%
     \includegraphics[width=0.45\textwidth]{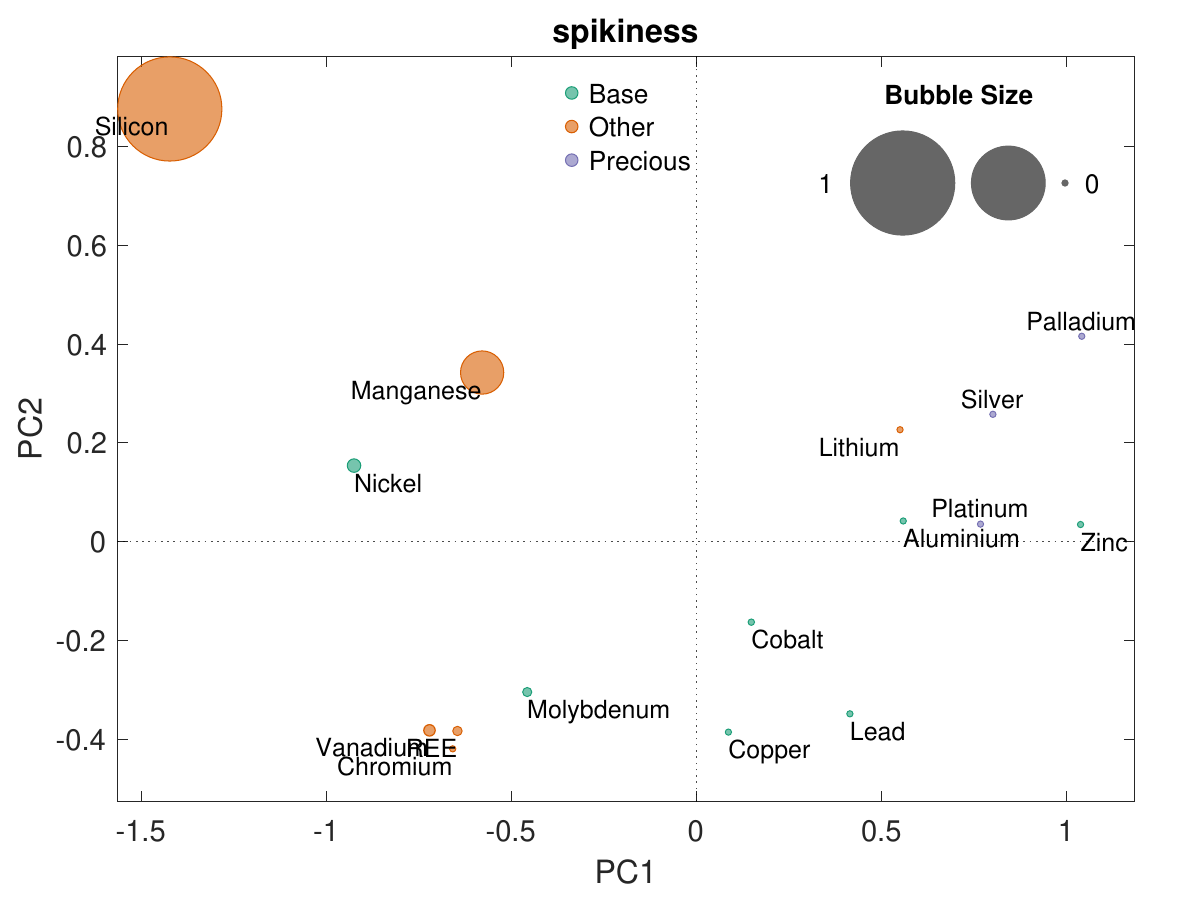}\\
    \label{fig:features2}
     \caption*{\scriptsize\textit{Notes}: each panel shows the instance space of RV time series. $PC_1$ and $PC_2$ are the first two principal components, projected from the eight-dimensional feature space. In each panel, the size of the dots is proportional to the value of the feature in the title. Features are normalized in the (0,1) interval.}
\end{figure}

\FloatBarrier 

As far as skewness, kurtosis and entropy are concerned, the graphs in panels 4, 5 and 7 show that nickel is associated with values of these features that are much higher than those of most other base metals, which explains why it tends to be far from them in the instance space. Furthermore, we can see int Figure \ref{fig:featheatmap} tha nickel is associated with very low values for the ACF features. As for silicon, we can see that it is associated with low values of ACF features and high values of distributional and other features, especially spikiness, which explains why it tends to appear as an outlier in the top-left corner of the instance space.

\section{Evaluation of volatility models and forecasts}\label{sec:results2}
\subsection{Model comparison}\label{sec:modelcomp}
The in-sample performance of the $m$-th volatility model, $\mathcal{M}_m$, can be assessed focusing on its marginal likelihood (ML), defined as: 
\begin{equation}
\label{eq:ml}
p\left( \mathbf{r} \left| \mathcal{M}_m \right. \right) =%
\int  p\left(\mathbf{r} \left| \boldsymbol{\theta}_m,  \mathcal{M}_m \right. \right) %
p\left(\boldsymbol{\theta}_m \left| \mathcal{M}_m \right. \right) {d} \boldsymbol{\theta}_m
\end{equation}
where $ p\left(\mathbf{r} \left| \boldsymbol{\theta}_m,  \mathcal{M}_m \right. \right)$ is the likelihood function and the prior density $p\left(\boldsymbol{\theta}_m \left| \mathcal{M}_m \right. \right)$, both of which depend on a model-specific vector of parameters, $\boldsymbol{\theta}_m$, of size $p_m$. 
The ML captures the fact that if data, collected in the vector of returns $\mathbf{r}$, are likely under model $m$, $p\left( \mathbf{r} \left| \mathcal{M}_m \right. \right)$ should be large. In other terms, the ML is akin to a density forecast of the returns under the $k$-th model, evaluated at the observed data. Moreover, we note that since the ML depends on the vector of parameters $\boldsymbol{\theta}_m$, it discounts model complexity from fit.

Denoting the GARCH(1,1) model as $\mathcal{M}_0$ we can compare it against SV specifications using the Bayes factor defined as: $BF_{m,0}=\frac{p\left( \mathbf{r} \left| \mathcal{M}_m \right. \right)}{p\left( \mathbf{r} \left| \mathcal{M}_0 \right. \right)}$ for $k = 1,...,5$. Therefore, observed returns are more likely under the $m$-th SV specification, compared to the GARCH(1,1) model, $\mathcal{M}_0$, if $BF_{i,0}>1$. As shown in \citet{kass1995bayes}, the log $BF$ is asymptotically equivalent to the Schwarz information criterion; moreover, both are consistent model selection criteria (i.e. they can asymptotically recover the model with the correct structure with probability one).

\begin{table}[!ht]
\newcolumntype{Z}{>{\small\raggedright\arraybackslash}X}
\newcolumntype{Y}{>{\small\centering\arraybackslash}X}
    \centering
    \caption{Log marginal likelihood of the volatility models}
    \label{tab:ml}
   \begin{tabularx}{\textwidth}{ZYYYYYY}\hline
Commodity & GARCH(1,1) & SV(1) & SV(2) & SV(1)-J & SV(1)-t & SV(1)-L \\\hline
Aluminum & -366.2 & -366.5 & -366.1 & -367.2 & -366.3 & \textbf{-365.9} \\
         & (0.03) & (0.11) & (0.03) & (0.11) & (0.01) & (0.02) \\
Cobalt   & -443.2 & -445.1 & \textbf{-438.2} & -445.8 & -440.2 & -443.9 \\
         & (0.02) & (0.31) & (0.02) & (0.19) & (0.03) & (0.02) \\
Copper   & -374.8 & -372.9 & \textbf{-370.3} & -372.6 & -371.4 & -373.2 \\
         & (0.03) & (0.08) & (0.01) & (0.11) & (0.02) & (0.02) \\
Lead     & -376.8 & -376.6 & \textbf{-373.6} & -377.0 & -376.7 & -376.9 \\
         & (0.11) & (0.10) & (0.01) & (0.09) & (0.01) & (0.02) \\
Molybdenum & -464.7 & -444.8 & \textbf{-427.3} & -447.3 & -427.5 & -439.1 \\
         & (0.07) & (0.28) & (0.06) & (0.42) & (0.02) & (0.08) \\
Nickel   & -438.4 & -434.0 & -432.0 & -434.3 & \textbf{-431.7} & -432.0 \\
         & (0.04) & (0.17) & (0.02) & (0.26) & (0.01) & (0.01) \\
Zinc     & -400.7 & -401.3 & \textbf{-399.8} & -401.4 & -400.3 & -401.2 \\
         & (0.02) & (0.09) & (0.02) & (0.14) & (0.01) & (0.01) \\\hline
Silver   & \textbf{-416.8} & -420.3 & -418.6 & -420.5 & -418.4 & -419.3 \\
         & (0.02) & (0.10) & (0.02) & (0.12) & (0.01) & (0.01) \\
Palladium & -391.2 & -388.0 & \textbf{-386.7} & -388.5 & -387.2 & -387.6 \\
         & (0.08) & (0.11) & (0.02) & (0.10) & (0.01) & (0.01) \\
Platinum & -410.4 & -409.1 & \textbf{-407.4} & -409.6 & -407.9 & -408.8 \\
         & (0.03) & (0.12) & (0.03) & (0.18) & (0.01) & (0.01) \\\hline
Chromium & -369.9 & -363.2 & \textbf{-359.0} & -363.0 & -360.0 & -363.3 \\
         & (0.08) & (0.22) & (0.02) & (0.30) & (0.06) & (0.02) \\
Lithium  & -369.0 & -325.7 & -323.9 & -325.8 & \textbf{-323.6} & -325.4 \\
         & (0.12) & (0.13) & (0.05) & (0.16) & (0.07) & (0.03) \\
Manganese & -453.3 & -436.3 & \textbf{-424.5} & -434.4 & -428.0 & -432.5 \\
         & (0.04) & (0.28) & (0.04) & (0.33) & (0.03) & (0.02) \\
REE      & -417.4 & -362.5 & -362.2 & -363.7 & \textbf{-358.6} & -359.9 \\
         & (0.02) & (0.20) & (0.49) & (0.25) & (0.05) & (0.08) \\
Silicon  & -449.9 & -346.0 & \textbf{-328.2} & -349.2 & -340.7 & -343.4 \\
         & (0.08) & (0.03) & (0.22) & (0.43) & (0.16) & (0.17) \\
Vanadium & -482.7 & -397.7 & \textbf{-374.1} & -399.3 & -387.0 & -390.3 \\
         & (0.05) & (0.36) & (0.26) & (0.36) & (0.17) & (0.22) \\\hline

\end{tabularx}
\caption*{\scriptsize\textit{Notes}: values in bold represent the best model associated with the highest marginal likelihood. Numerical standard errors are given in brackets.}
\end{table}

\FloatBarrier

\begin{figure}
    \centering
    \caption{Instance space and Bayes Factors}
    \includegraphics[width=0.495\linewidth]{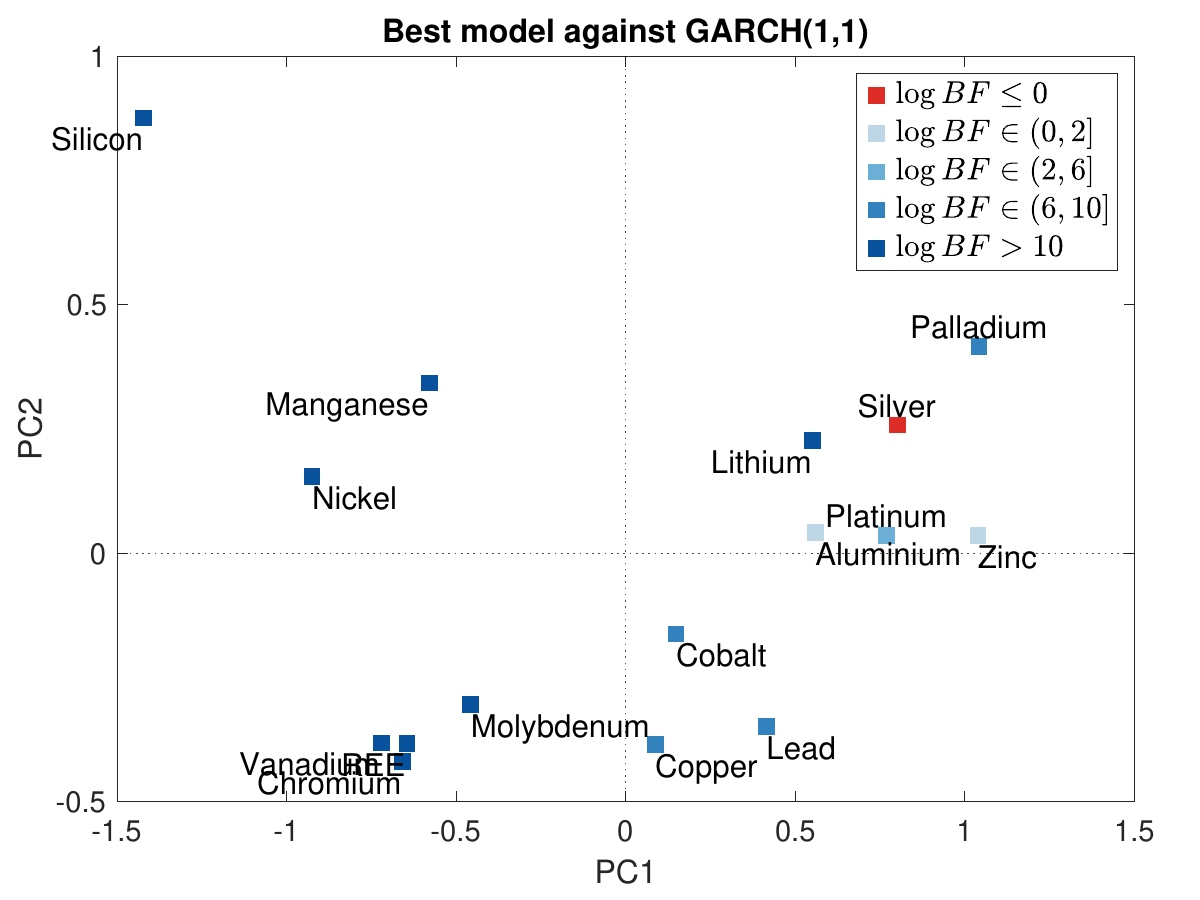}%
    \includegraphics[width=0.495\linewidth]{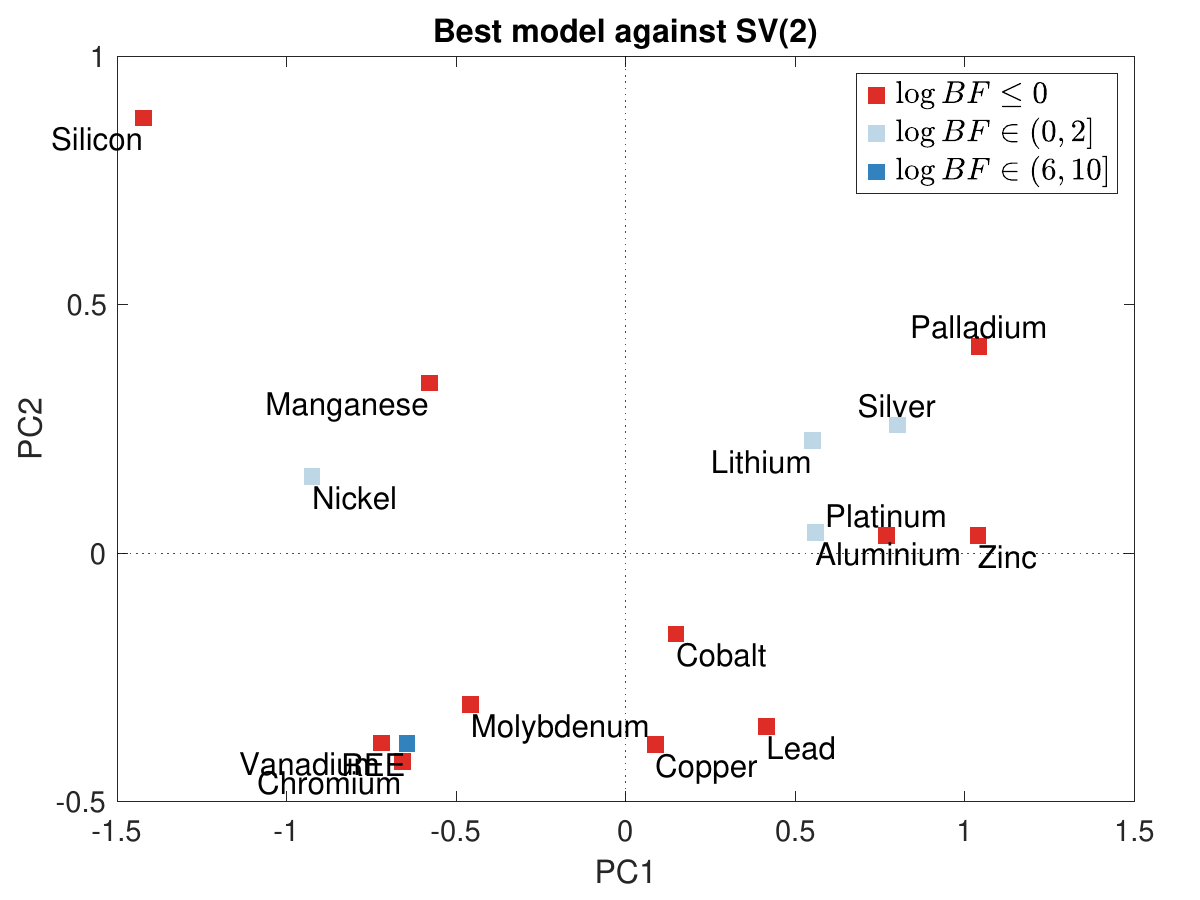}
    \label{fig:figlogml}
\caption*{\scriptsize\textit{Notes}: the colour of the dots is based on the scale given by \citet{kass1995bayes} for interpreting twice the natural logarithm of the Bayes factor. A negative value (red dot) provides no evidence in favour of the best model against the GARCH(1,1) or the SV(2) model in the left and right panels, respectively. Dots on a blue scale indicate positive values and provide evidence (weak, positive, strong, very strong) in favour of the best model compared to one of the two benchmarks. See notes to Figure \ref{fig:features2} for further details. }
\end{figure}

Table \ref{tab:ml} shows the log ML across metals and models. As we can see from the first column of the table, silver is the only metal for which the log ML provides evidence in favor of the GARCH(1,1) against SV specifications. This is also clear from the left panel of Figure \ref{fig:figlogml}, which plots (twice the log of) the BF of the best model (i.e. the one associated with the largest or second-largest log ML) against the GARCH(1,1) model. Silver is associated with a negative value of log BF, shown as a red dot, indicating that there is no evidence to support any of the SV specifications against a simple GARCH(1,1) model.

Conversely, the dots representing the other metals are on a blue scale, indicating positive values of log BF. We can also see that the further we move away from the horizontal zero line, the greater the log BF (darker dots). Recall that in Section \ref{sec:feat} we showed that $PC_2$ captures distributional features, spikiness, and zero returns, while $PC_1$ captures mostly ACF features. Therefore, as we move away from the horizontal zero line, the contribution of $PC_2$ to the variance of the data increases. As this happens, the performance of the GARCH(1,1) model, in which conditional volatility is a deterministic function of past data and can therefore capture mostly ACF features, deteriorates. In these cases, the additional flexibility of SV models, where the log-volatility equation includes a random shock, seems to provide some gains.

The fourth column of Table \ref{tab:ml} highlights that the SV(2) model provides the best fit for most metals. This is also clear from the right panel of Figure \ref{fig:figlogml}, which plots (twice the log of) the BF of the best model (i.e. the one associated with the largest or second-largest log ML) against the SV(2) model. Note that most dots are red and hence indicate a negative log BF, which represent evidence in favor of the SV(2) model.

However, the plot also highlights metals for which there is no evidence in favor of the SV(2) model: aluminum, nickel, silver, lithium, and REE. For aluminum, nickel, silver, and lithium, the dots are in light shades of blue to indicate that $\log BF \in (0,2]$ and hence there is only weak evidence in favor of alternative specifications of the SV relative to the SV(2) model. Note that the dots associated with these metals lie near the horizontal zero line in the instance space, thus suggesting an increasing contribution of $PC_2$ to explaining the variance of the data. Only for REE there is strong evidence that an alternative SV model provides gains over the SV(2). Interestingly, looking at Table \ref{tab:ml}, we can see that for nickel, lithium, and REE, the SV(1)-$t$ specification performs well. While these metals lie in different regions of the instance space, Figure \ref{fig:featheatmap} shows that they exhibit strong values for entropy, skewness, and kurtosis.

Overall, the results in this section point to the SV(2) as the model that best accommodates the heterogeneous features of ETM data. The presence of a second autoregressive log-volatility term seems, therefore, to capture the set of complex patterns that characterizes the ETM time series.

\subsection{Point forecasts}\label{sec:pointfore}
As outlined in Section \ref{sec:foreex}, models are recursively estimated to produce a total of 60 one-step-ahead forecasts from January 2018 to December 2022. Point volatility forecasts are evaluated against two volatility proxies -- RV and $RG2^*$ -- using two alternative loss functions: the RMSFE and the QLIKE. Table \ref{tab:RVpointfore} focuses on RV as a volatility proxy, while results for RG2* appears in Appendix \ref{sec:appfigtab}. We implement the test for equal predictive ability of \cite{diebold1995comparing}, as modified by \cite{coroneo2020comparing} to improve its performance in small samples (DM test, henceforth). The null hypothesis of the test is that the observed difference in forecasting performance of SV specifications against the GARCH(1,1) model is statistically indistinguishable from zero.

Each entry in Table \ref{tab:RVpointfore} represents (if multiplied by 100) the percentage change in loss when moving from a GARCH(1,1) forecast to one based on an SV specification. For instance, in the case of aluminum, the SV(1) forecast results in a 1.76\% reduction in RMSFE and a 28.43\% decrease in QLIKE loss.

\begin{table}[ht]
\newcolumntype{Z}{>{\small\raggedright\arraybackslash}X}
\newcolumntype{Y}{>{\small\centering\arraybackslash}X}
    \centering
    \caption{Evaluation of volatility point forecasts using RV as proxy}
    \label{tab:RVpointfore}
    \begin{tabularx}{\textwidth}{ZYYYYY}
\hline
\multicolumn{6}{c}{\small (\textit{a}) Root Mean Squared Forecast Error loss}\\\hline
     & SV(1) & SV(2) & SV(1)-J & SV(1)-t & SV(1)-L \\
\hline
    Aluminum    & {-0.0176}$^{*}$ & 0.0587      & -0.0101     & -0.0078     & \textbf{-0.0182}     \\
    Cobalt      & 0.1293        & 0.1400      & 0.1393      & 0.1617      & 0.1322      \\
    Copper      & \textbf{-0.0094}       & 0.0335      & 0.0204      & 0.0259      & -0.0053     \\
    Lead        & -0.0719$^{**}$ & -0.0345$^{**}$ & -0.0515$^{**}$ & -0.0500$^{**}$ & \textbf{-0.0777}$^{**}$ \\
    Molybdenum  & 1.8386        & 0.2244      & 0.0313      & 0.0466      & 0.0637      \\
    Nickel      & -0.0026       & 0.0011      & -0.0023     & -0.0004     & \textbf{-0.0033}$^{*}$ \\
    Zinc        & 0.0020        & 0.0889      & 0.0233      & 0.0363      & 0.0127      \\\hline
    Silver      & 0.0421        & 0.1054      & 0.0451      & 0.0511      & 0.0463      \\
    Palladium   & \textbf{-0.0416}$^{**}$ & 0.0215      & -0.0153$^{**}$ & -0.0122$^{**}$ & -0.0355$^{**}$ \\
    Platinum    & -0.0930$^{**}$ & -0.0173     & -0.0809$^{**}$ & -0.0679$^{**}$ & \textbf{-0.0967}$^{**}$ \\\hline
    Chromium    & 0.0885        & \textbf{-0.0370}$^{**}$ & -0.0291     & -0.0012     & 0.1446      \\
    Lithium     & 0.4439        & -0.1019     & 0.1646      & \textbf{-0.1581}     & 0.5451      \\
    Manganese   & -0.0014       & \textbf{-0.0126}     & -0.0041     & -0.0133     & -0.0070     \\
    REE         & 0.3698        & 0.2324      & 0.2060      & 0.1928      & 0.2025      \\
    Silicon     & 0.2381        & 0.0257      & 0.2818      & 0.7451      & \textbf{-0.0063}     \\
    Vanadium    & 0.4387        & 0.2529      & 0.5309      & 0.5881      & 0.0398      \\\hline
\multicolumn{6}{c}{\small (\textit{b}) QLIKE loss}\\\hline
     & SV(1) & SV(2) & SV(1)-J & SV(1)-t & SV(1)-L  \\\hline
    Aluminum    & -0.2843$^{**}$ & -0.0112     & -0.2274$^{*}$ & -0.2296$^{*}$ & \textbf{-0.3181}$^{**}$ \\
    Cobalt      & 0.0738         & 0.0546      & 0.1140        & 0.2164        & 0.0920         \\
    Copper      & 0.0186         & 0.2493      & 0.2599        & 0.2843        & 0.0834         \\
    Lead        & -0.3279$^{**}$ & -0.1866$^{**}$ & -0.2398$^{**}$ & -0.2358$^{**}$ & \textbf{-0.3402}$^{**}$ \\
    Molybdenum  & 0.1134         & 0.2386      & 0.7838        & 0.7711        & 0.2323         \\
    Nickel      & 0.0063         & 0.0159      & 0.0353        & 0.0716        & \textbf{-0.1338}        \\
    Zinc        & 0.0998         & 0.2252      & 0.2082        & 0.2545        & 0.2248         \\
\hline
    Silver      & \textbf{-0.0739}        & 0.5476      & -0.0409       & 0.0082        & 0.0181         \\
    Palladium   & \textbf{-0.2430}$^{**}$ & 0.0856      & -0.1187$^{**}$ & -0.1151$^{*}$ & -0.2076$^{**}$ \\
    Platinum    & \textbf{-0.3906}$^{**}$ & -0.2063     & -0.3477$^{**}$ & -0.3203$^{**}$ & -0.3562$^{**}$ \\
\hline
    Chromium    & 0.1021         & -0.0399     & 0.0263        & 0.2800        & 0.1114         \\
    Lithium     & 0.2764         & -0.0119     & 0.0279        & \textbf{-0.0585}       & 0.1684         \\
    Manganese   & 0.0799         & 0.5965      & 0.2557        & 0.5985        & 0.1164         \\
    REE         & -0.1586        & \textbf{-0.1911}     & 0.2420        & 0.3806        & -0.1279        \\
    Silicon     & -0.1849        & 0.2518      & -0.0963       & \textbf{-0.2449}       & 0.4818         \\
    Vanadium    & 1.5085         & 0.6045      & 1.4553        & 1.5907        & 3.4599         \\
\hline
    
    \end{tabularx}
    \caption*{\scriptsize\textit{Notes}: each entry in panel (\textit{a}) of the table shows $(RMSFE_j/RMSFE_0) - 1$, where $RMSFE_0$ is the root mean squared forecast error of the $GARCH(1,1)$ forecast, and $RMSFE_j$ corresponds to one of the SV specifications: $j = SV(1), SV(2), SV(1)-J, SV(1)-t, SV(1)-L$. In panel (\textit{b}) the loss function is the QLIKE. A negative value indicates that the $GARCH(1,1)$ forecast is less accurate than that of the corresponding SV specification. Asterisks denote the rejection of the null hypothesis of equal predictive ability in the Diebold-Mariano test, with $^{*}$ indicating significance at the 10\% level and $^{**}$ at the 5\% level.}
\end{table}
\FloatBarrier

The large difference between these figures is explained by the fact that, while the RMSFE is a symmetric loss function, the QLIKE loss penalizes underforecasting more heavily than overforecasting. In both cases, the table displays an asterisk indicating that, according to the DM test, we can reject the null hypothesis that the observed difference in forecasting performance between the SV(1) model and the GARCH(1,1) model is statistically indistinguishable from zero.

Although there is considerable heterogeneity in the accuracy of SV forecasts across different metal categories, some common patterns emerge. Looking at Table \ref{tab:RVpointfore}, we can see that in a total of 160 pairwise comparisons with GARCH(1,1) forecasts, SV specifications yield a reduction in the loss function in 41.3\% of cases, and we can reject the null of equal predictive accuracy in 20\% of cases.

The reductions in the loss function associated with the SV models vary across mineral classes and specifications. Focusing on base metals, we observe that SV forecasts are more accurate than GARCH(1,1) forecasts for aluminum, lead, copper, and nickel. However, for the latter two metals, the reductions in forecast errors are often statistically indistinguishable from zero. For the precious metals, the SV specifications lead to an improvement in forecast accuracy for palladium and platinum, but not for silver. This is in line with the model comparison in Section \ref{sec:modelcomp} where we showed that silver was the only metal for which none of the SV specifications performed well against the GARCH(1,1) model. Turning to the ``other ETMs'', we observe reductions in the loss function associated with the SV models in 20 out of 60 pairwise comparisons, but only in one case can we reject the null hypothesis of the DM test.

\begin{figure}[!ht]
    \caption{Instance space and volatility point forecasts}
    \centering
    \includegraphics[width=0.495\linewidth]{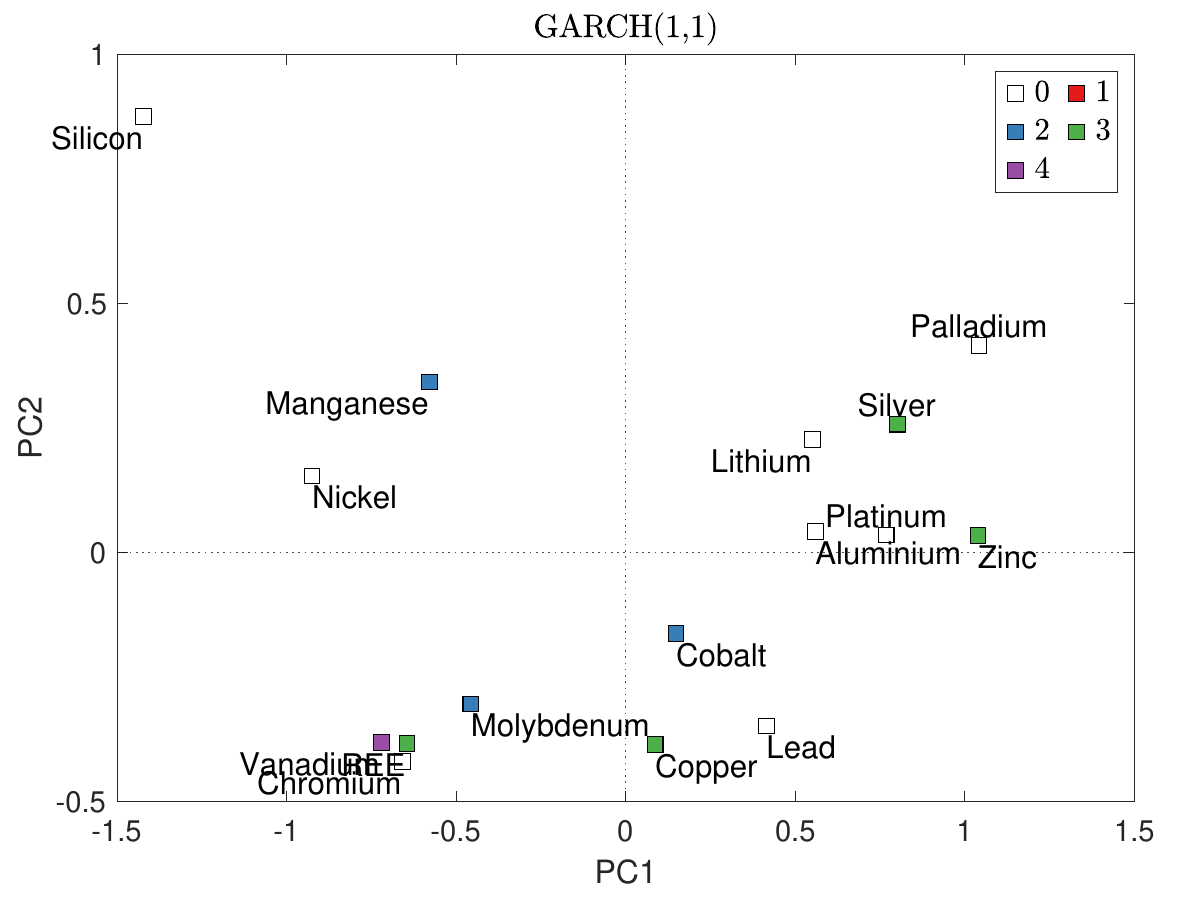}%
    \includegraphics[width=0.495\linewidth]{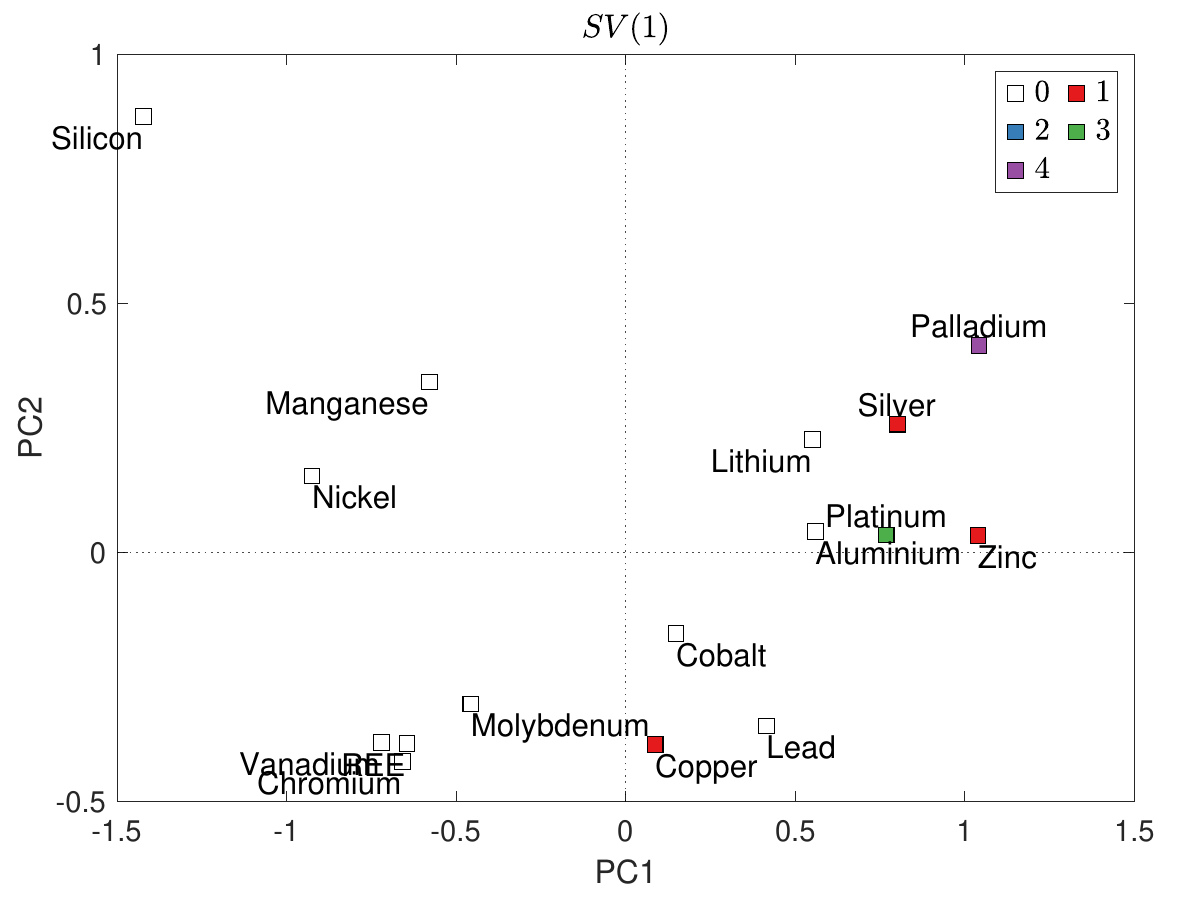}\\
    \includegraphics[width=0.495\linewidth]{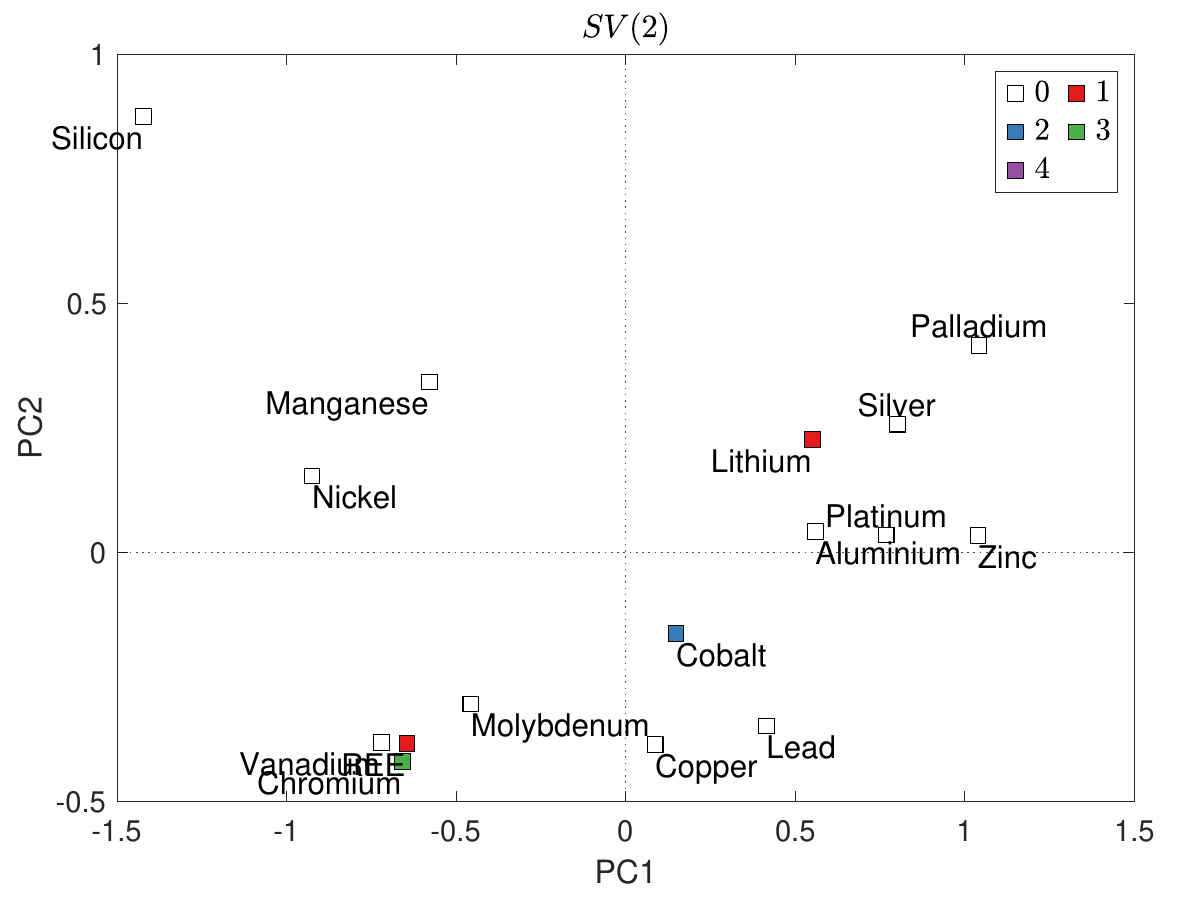}%
    \includegraphics[width=0.495\linewidth]{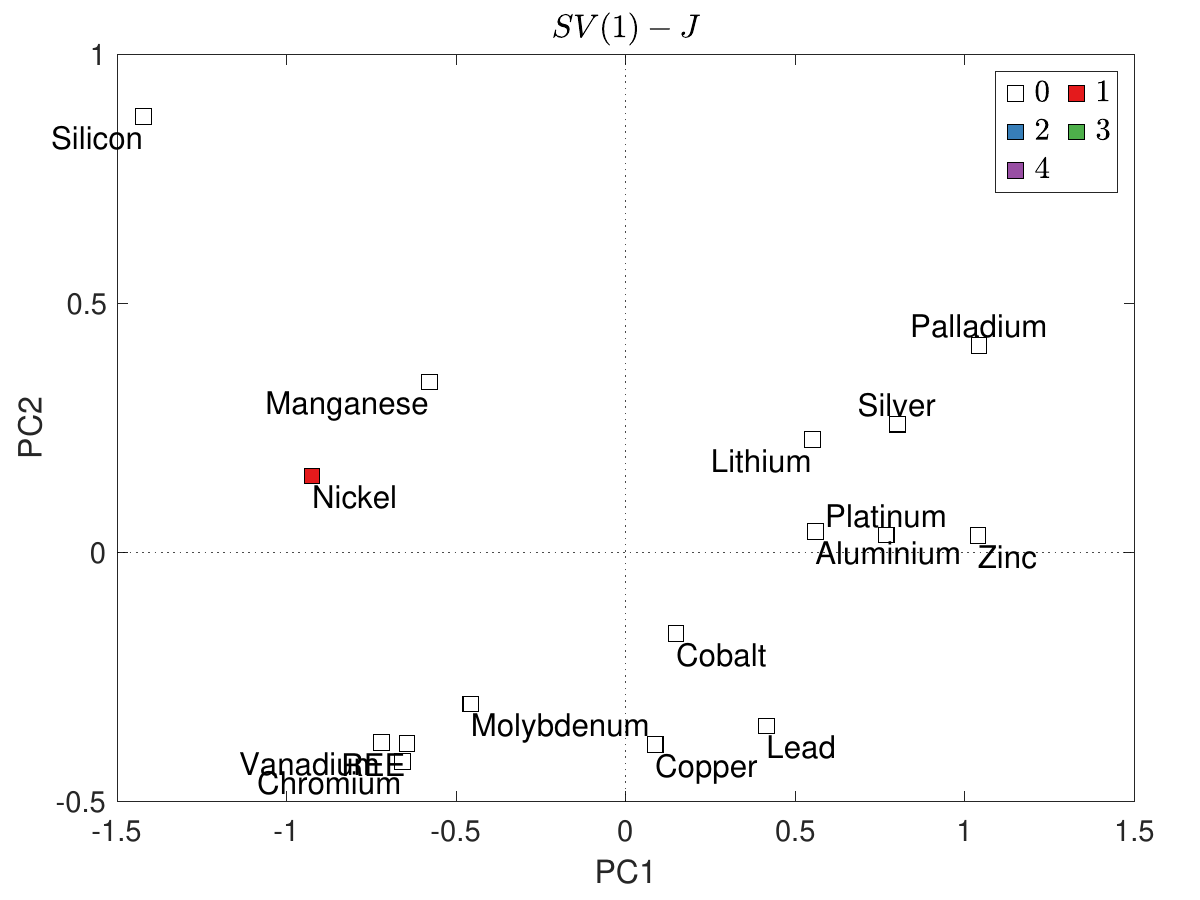}\\
        \includegraphics[width=0.495\linewidth]{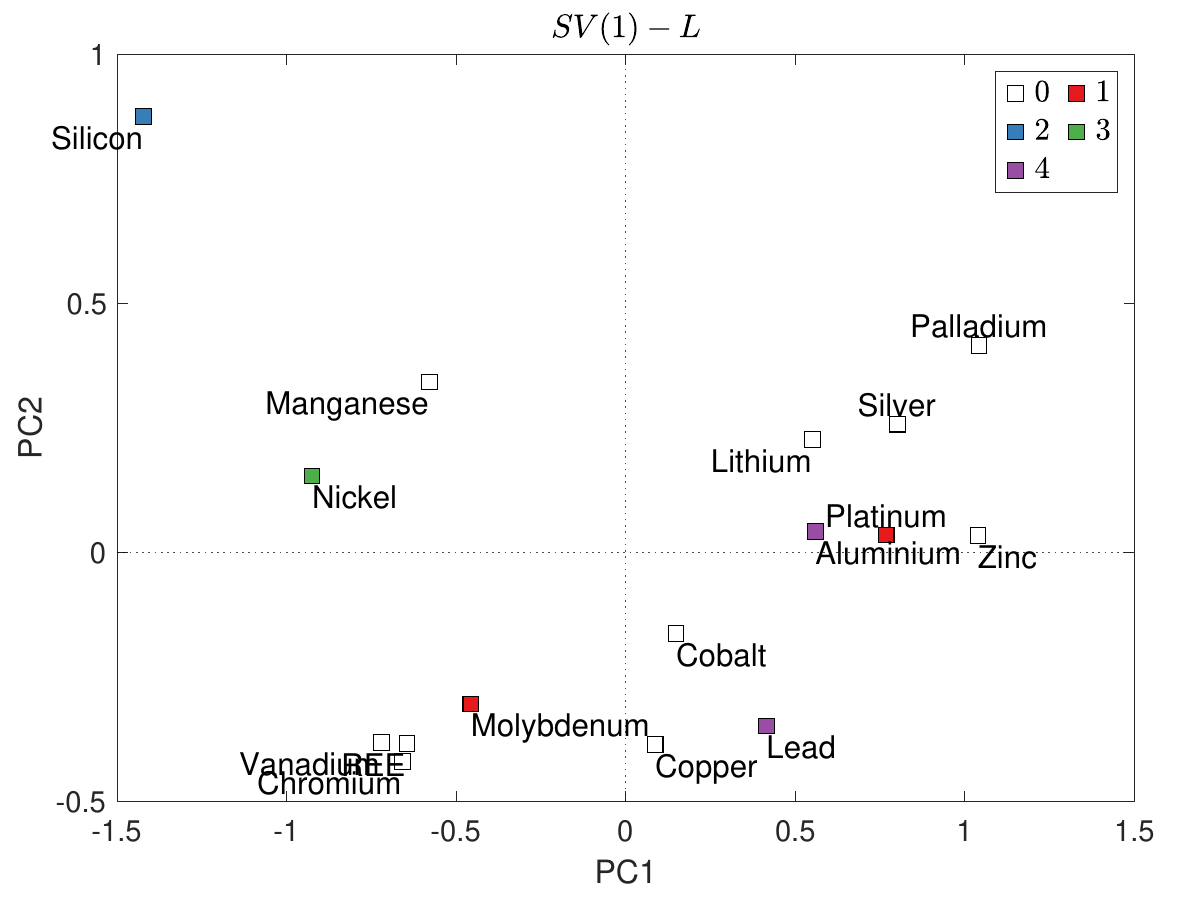}%
    \includegraphics[width=0.495\linewidth]{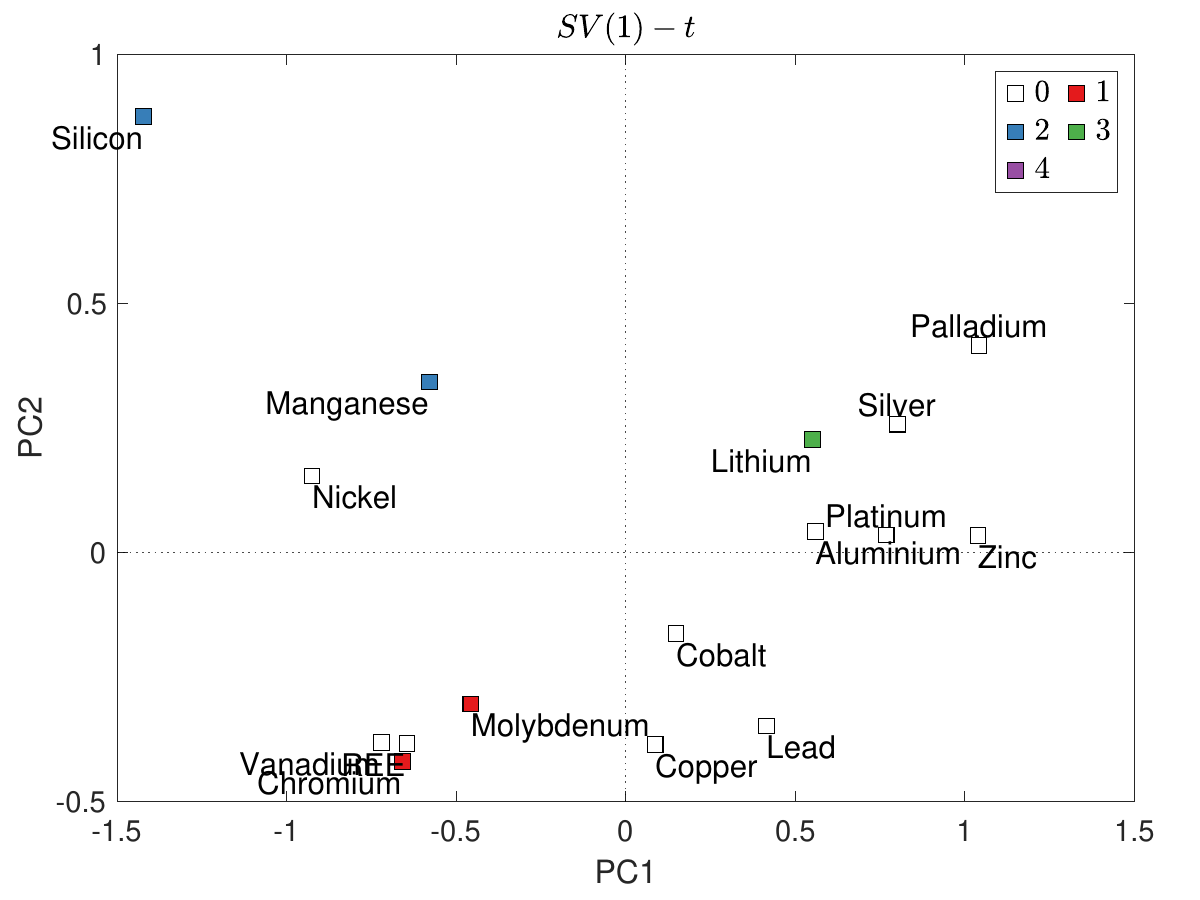}
    \label{fig:figFeatloss}
\caption*{\scriptsize\textit{Notes}: the colours of the dots indicate the number of times that the model in the title yields the lowest loss for a given metal. For each model we consider 2 loss functions and 2 volatility proxies, so a model can be selected a maximum of four times for each metal. See notes to Figure \ref{fig:features2} for further details.}
\end{figure}

As shown in Table \ref{tab:RVpointfore} and Figure \ref{fig:figFeatloss}, which also incorporates results based on the RG2* volatility proxy, the SV(1)-J forecasts are rarely more accurate than those of the GARCH(1,1) model. On the contrary, the specification that most often appears to offer the largest percentage change loss reduction compared with GARCH(1,1) forecasts is the SV(1)-L model.

A closer look at Figure \ref{fig:figFeatloss}, where in each panel we show the instance space and highlight the number of times that the model yields the lowest loss for a given metal, reveals that often the most accurate forecasts are either those from SV(1) and SV(2) models or those from the SV(1)-L. This pattern is confirmed across different classes of metals and seems to be related both to ACF and distributional features.

Three main conclusions can be drawn from the analysis in this section. First, contrary to the in-sample model comparison in Section \ref{sec:modelcomp}, where the SV(2) model appeared to provide the best fit to the ETM data, there seems to be no clear winner in the case of point volatility forecasts. This, in turn, suggests that forecast combination might be a suitable approach for improving point volatility forecasts. Second, in the case of point forecasts, accounting for the asymmetric effects of positive and negative returns on volatility might lead to more accurate predictions. Since the SV(1)-L model rarely outperforms the SV(2) specification when focusing on the log BF, this underscores the importance of not relying solely on in-sample metrics for selecting models to produce volatility point forecasts. Lastly, there are no clear patterns linking point forecast accuracy to the prevalence of specific features.

\FloatBarrier

\subsection{Density Forecasts}\label{sec:densfore}
For density forecasts in Table \ref{tab:DensityFore}, we focus on weighted versions of the qCRPS that emphasize either the center (Panel \textit{a}) or both tails of the distribution of returns (Panel \textit{b}). The structure and interpretation of the entries in the table are the same as those in Table \ref{tab:RVpointfore}. Inspection of Table \ref{tab:DensityFore} reveals that, in a total of 160 pairwise comparisons with GARCH(1,1) forecasts, SV specifications yield a reduction in the loss function in 43.8\% of cases, and the null of equal predictive accuracy can be rejected in 23.1\% of cases. Furthermore, the most accurate forecast (highlighted in bold) is often associated with the SV(2) specification.

\begin{figure}[!ht]
    \caption{Instance space and density forecasts}
    \centering
    \includegraphics[width=0.495\linewidth]{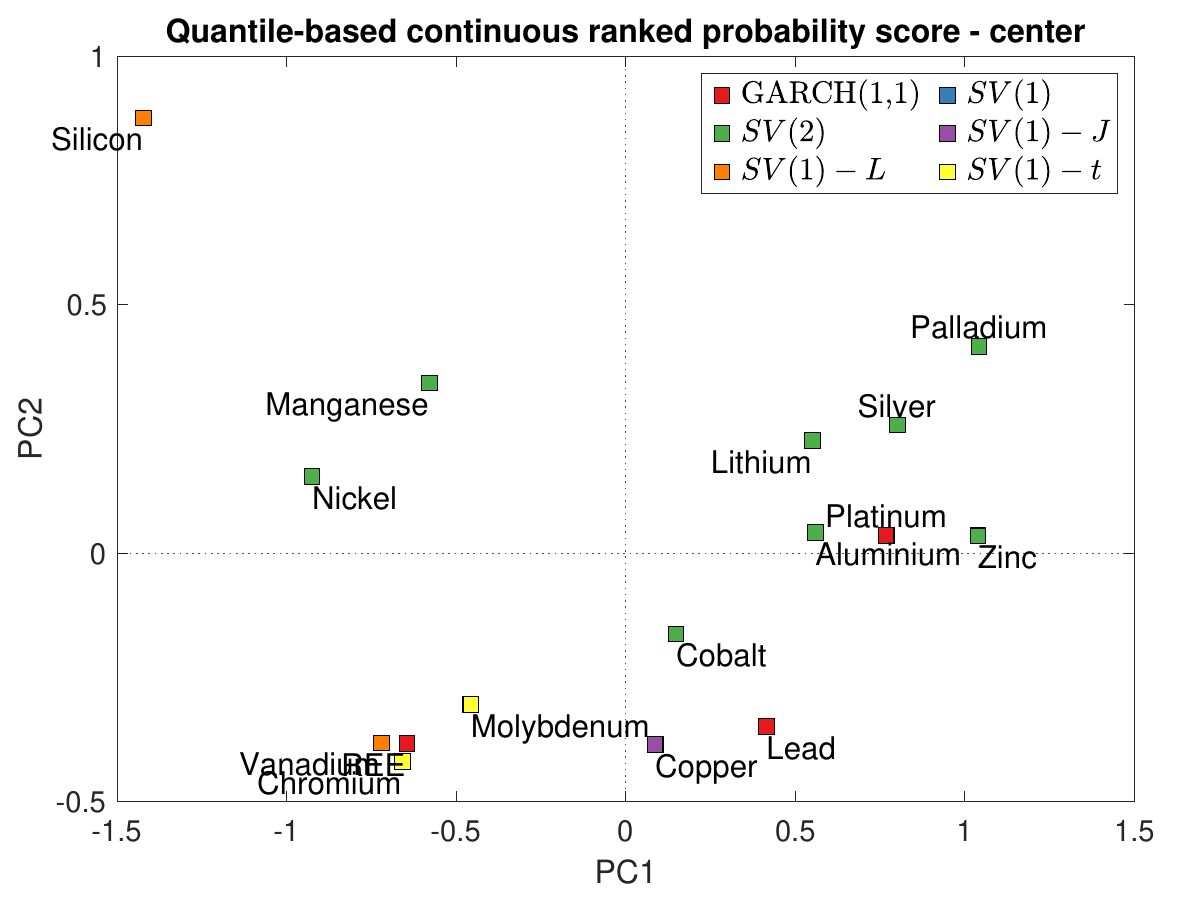}%
    \includegraphics[width=0.495\linewidth]{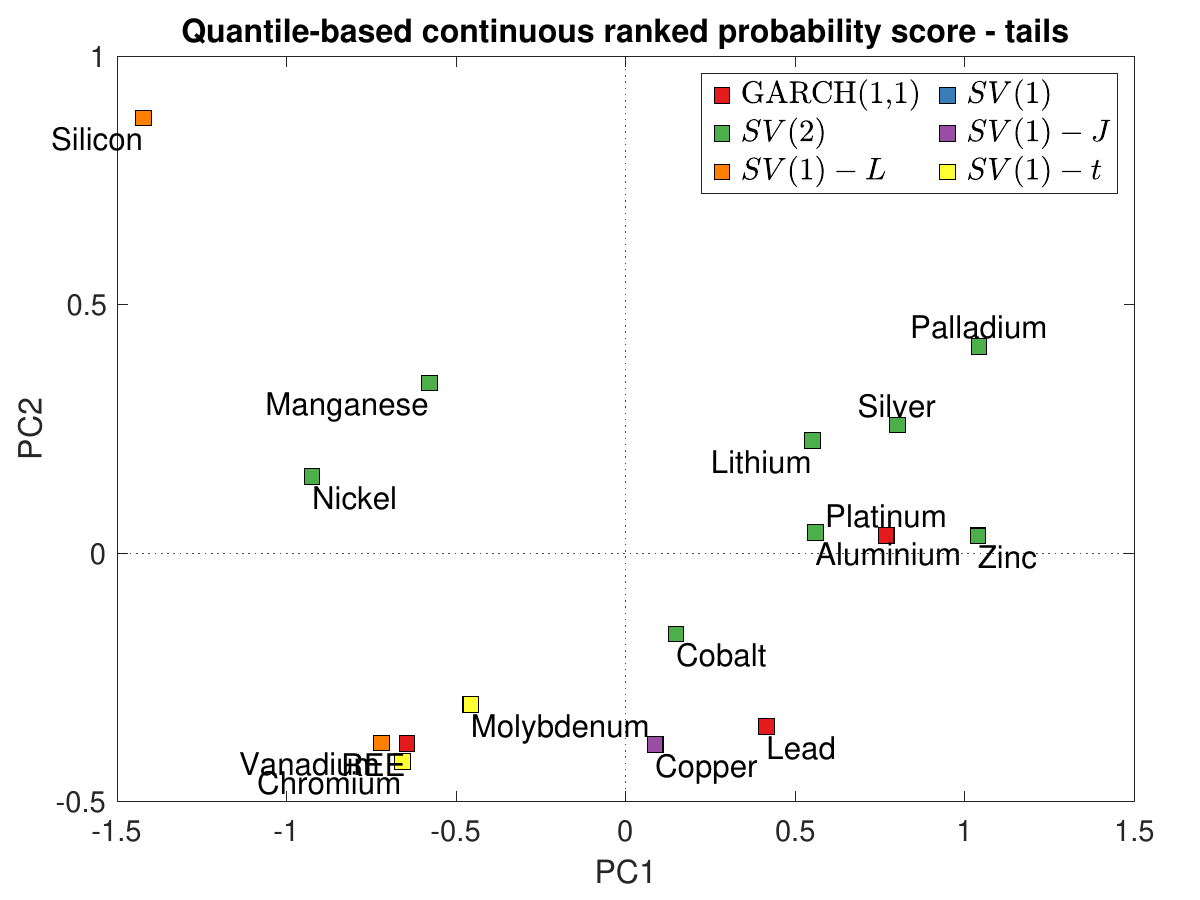}%
    \label{fig:figQS}
\caption*{\scriptsize\textit{Notes}: The colours of the dots indicate which model delivers the lowest loss. The loss function is given in the title. See notes to Figure \ref{fig:features2} for further details.}
\end{figure}

In Figure \ref{fig:figQS}, we indicate the most accurate forecast for each metal represented in the instance space. For metals located near the horizontal zero line, the best forecasts are typically associated with models such as SV(2) and GARCH(1,1), which emphasize the correlation in volatilities to produce density forecasts. As we move farther from the horizontal line, the importance of distributional features, such as spikiness and entropy, in explaining the variability of the data increases. In these cases, models like SV(1)-L and SV-t are often associated with the most accurate density forecasts.

\begin{table}[!ht]
    \newcolumntype{Z}{>{\small\raggedright\arraybackslash}X}
\newcolumntype{Y}{>{\small\centering\arraybackslash}X}
    \centering
    \caption{Quantile-based continuous ranked probability score (qCRPS)}
    \label{tab:DensityFore}
    \begin{tabularx}{\textwidth}{ZYYYYY}
\hline
\multicolumn{6}{c}{\small (\textit{a}) Weighted qCRPS ($\widehat{wQS}$) - center}\\\hline
     & SV(1) & SV(2) & SV(1)-J & SV(1)-t & SV(1)-L \\
\hline
    Aluminum    & 0.0334        & \textbf{-0.0129 }    & 0.0201      & 0.0243       & 0.0216      \\
    Cobalt      & 0.0099        & \textbf{-0.0801}$^{**}$ & -0.0085     & -0.0738$^{**}$ & 0.0138      \\
    Copper      & 0.0062        & -0.0255$^{**}$ & \textbf{-0.0305}$^{**}$ & -0.0274$^{**}$ & -0.0030     \\
    Lead        & 0.0600        & 0.0320 & 0.0428 & 0.0468 & 0.0635 \\
    Molybdenum  & 0.4082        & -0.0961$^{**}$ & -0.2229$^{**}$ & \textbf{-0.3018}$^{**}$ & -0.1393$^{**}$ \\
    Nickel      & 0.0206        & \textbf{-0.0291}     & 0.0092      & -0.0030     & 0.0232       \\
    Zinc        & -0.0044       & \textbf{-0.0405}$^{**}$ & -0.0153     & -0.0167     & -0.0049     \\
\hline
    Silver      & -0.0139       & \textbf{-0.0786}$^{**}$ & -0.0203     & -0.0238     & -0.0227     \\
    Palladium   & 0.0412        & \textbf{-0.0167}$^{*}$ & 0.0123       & 0.0157      & 0.0331       \\
    Platinum    & 0.0468        & 0.0102      & 0.0275      & 0.0222      & 0.0524      \\
\hline
    Chromium    & 0.0389        & -0.0479$^{**}$ & -0.0739$^{**}$ & \textbf{-0.1342}$^{**}$ & 0.0511      \\
    Lithium     & 0.1783        & -0.0032     & 0.0700      & 0.0008      & 0.1572       \\
    Manganese   & 0.0211        & \textbf{-0.1966}$^{**}$ & -0.0294     & -0.1735$^{**}$ & 0.0134      \\
    REE         & 0.3375        & 0.2987      & 0.2216      & 0.2238      & 0.2579       \\
    Silicon     & 0.4779        & 0.0042      & 0.3462       & 0.6023       & \textbf{-0.1098 }    \\
    Vanadium    & 0.0676        & -0.0402     & 0.1310      & 0.1865      & \textbf{-0.2583}$^{**}$ \\
\hline
\multicolumn{6}{c}{\small (\textit{b})  Weighted qCRPS ($\widehat{wQS}$) - tails}\\\hline
     & SV(1) & SV(2) & SV(1)-J & SV(1)-t & SV(1)-L \\
\hline

    Aluminum    & 0.0527 & -0.0256     & 0.0313      & 0.0412 & 0.0368      \\
    Cobalt      & 0.0124        & \textbf{-0.1192}$^{**}$ & -0.0174     & -0.1095$^{*}$ & 0.0174      \\
    Copper      & 0.0078        & -0.0358$^{**}$ & \textbf{-0.0454}$^{**}$ & -0.0369$^{**}$ & -0.0060     \\
    Lead        & 0.0924 & 0.0502 & 0.0657 & 0.0757 & 0.0972 \\
    Molybdenum  & 0.4914 & -0.1242$^{**}$ & -0.2808$^{**}$ & \textbf{-0.3405}$^{**}$ & -0.1797$^{**}$ \\
    Nickel      & 0.0344        & -0.0507     & 0.0166      & -0.0009     & 0.0366  \\
    Zinc        & -0.0059       & \textbf{-0.0635}$^{**}$ & -0.0232     & -0.0223     & -0.0079     \\
\hline
    Silver      & -0.0193       & \textbf{-0.1177}$^{**}$ & -0.0293     & -0.0313     & -0.0367     \\
    Palladium   & 0.0595 & \textbf{-0.0255}$^{**}$ & 0.0170  & 0.0251 & 0.0474 \\
    Platinum    & 0.0759  & 0.0214      & 0.0477      & 0.0435      & 0.0828  \\
\hline
    Chromium    & 0.0501        & -0.0726$^{**}$ & -0.1184$^{**}$ & \textbf{-0.1869}$^{**}$ & 0.0638      \\
    Lithium     & 0.2839 & -0.0087     & 0.1056 & 0.0024      & 0.2442\\
    Manganese   & 0.0327        & \textbf{-0.2738}$^{**}$ & -0.0480     & -0.2351$^{**}$ & 0.0271      \\
    REE         & 0.4621 & 0.4042& 0.3061& 0.3102 & 0.3572 \\
    Silicon     & 0.5506 & 0.0155      & 0.4001 & 0.7161  & \textbf{-0.1096 }    \\
    Vanadium    & 0.0900        & -0.0544     & 0.1779      & 0.2543  & \textbf{-0.3246}$^{**}$ \\
\hline
    \end{tabularx}
    \caption*{\scriptsize\textit{Notes}: each entry of the table shows $(\widehat{wQS}_j/\widehat{wQS}_0) - 1$, where $\widehat{wQS}_0$ is the weighted qCRPS the $GARCH(1,1)$ forecast, and $\widehat{wQS}_j$ corresponds to one of the SV specifications: $j = SV(1), SV(2), SV(1)-J, SV(1)-t, SV(1)-L$. In panel in panel (\textit{a}) the weighted qCRPS emphasizes the center of the distribution, while in panel (\textit{b}) the loss function emphasizes the tails of the distibution. A negative value indicates that the $GARCH(1,1)$ forecast is less accurate than that of the corresponding SV specification. Asterisks denote the rejection of the null hypothesis of equal predictive ability in the Diebold-Mariano test, with $^{*}$ indicating significance at the 10\% level and $^{**}$ at the 5\% level.}
\end{table}
\FloatBarrier
\FloatBarrier
\section{Conclusions}\label{sec:concl}
Some ETMs have liquid markets, while others have limited price histories and experience infrequent price changes. Modelling the volatilities of these commodities plays a crucial role in ensuring effective risk management during the energy transition process. The importance of modeling volatilities arises from the interplay of two critical factors: the increasing complexity of clean energy technologies and the low substitutability of ETMs. As a result, even minor shocks in ETM markets can trigger substantial price responses. A comprehensive understanding and evaluation of commodity volatilities is imperative for developing robust risk mitigation strategies.

 Our analysis, which combines exploratory data analysis, data reduction techniques, and visualization methods, reveals important insights into the volatility patterns of different ETMs. Despite the common classification of metals into homogeneous groups by data providers and their geological co-occurrence, we uncover significant heterogeneity in volatility patterns when focusing on key features of the time series of volatility and returns. This finding highlights the complexity of metal volatility dynamics. Furthermore, our evaluation of volatility models, both in-sample and out-of-sample, shows that neither GARCH models nor any single SV specification consistently outperforms others in forecasting experiments. This suggests that combining forecasts from different models or incorporating additional features may be necessary for improving volatility forecasts in the metal market. A case in point would be to explicitly incorporate zero returns in the estimation of models \citep{francq2022volatility}. We leave these topics for future research.

\newpage

\bibliography{BLS}

\newpage

\appendix
\section{Further tables and figures}\label{sec:appfigtab}
\FloatBarrier
\begin{figure}
    \centering
    \caption{Real prices of Energy transition metals}
    \includegraphics[width=\textwidth]{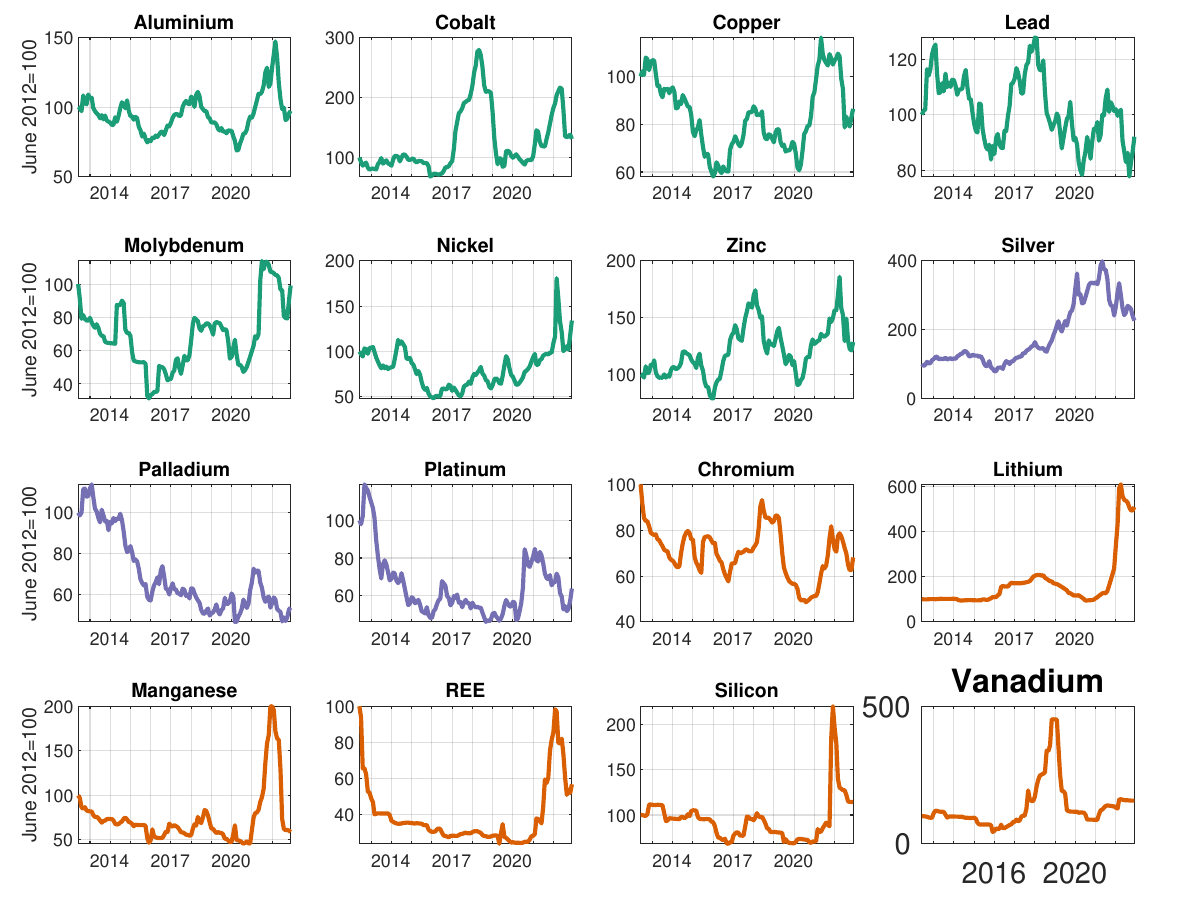}
    \label{fig:Pmetals}
\end{figure}

\begin{figure}
    \centering
    \caption{Realized Volatiltiy of Energy transition metals}
\includegraphics[width=\textwidth]{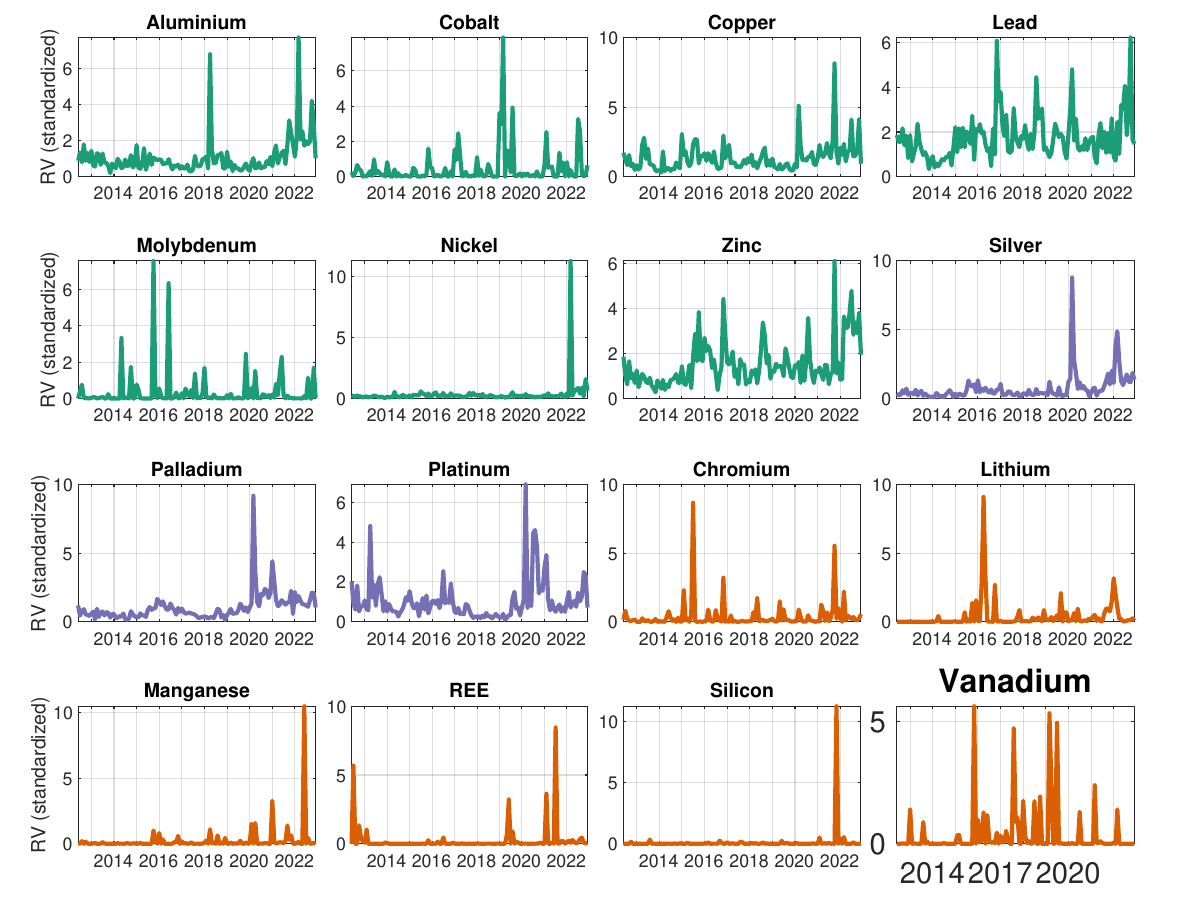}
    \label{fig:RVmetals}
\end{figure}

\begin{table}[!ht]
\newcolumntype{Z}{>{\small\raggedright\arraybackslash}X}
\newcolumntype{Y}{>{\small\centering\arraybackslash}X}
    \centering
    \caption{Evaluation of volatility point forecasts using RG2* as proxy}
    \label{tab:RG2pointfore}
    \begin{tabularx}{\textwidth}{ZYYYYY}
\hline
\multicolumn{6}{c}{\small (\textit{a}) Root Mean Squared Forecast Error loss}\\\hline
     & SV(1) & SV(2) & SV(1)-J & SV(1)-t & SV(1)-L \\
\hline
    Aluminum    & -0.0281 & -0.0104 & -0.0242 & -0.0249 & \textbf{-0.0513}$^{**}$ \\
    Cobalt      & -0.0802 & \textbf{-0.1756} & -0.0878 & -0.1450 & -0.0801 \\
    Copper      & 0.0053  & 0.0124  & 0.0118  & 0.0157  & 0.0116 \\
    Lead        & -0.0679$^{**}$ & -0.0322$^{**}$ & -0.0483$^{**}$ & -0.0469$^{**}$ & \textbf{-0.0694}$^{**}$ \\
    Molybdenum  & 3.5362  & 0.4073  & -0.1096 & \textbf{-0.1576}$^{*}$ & 0.0270 \\
    Nickel      & -0.0043 & 0.0025  & \textbf{-0.0060 }& -0.0034 & -0.0056 \\
    Zinc        & \textbf{-0.0210} & 0.0056  & -0.0128 & -0.0089 & -0.0131 \\
\hline
    Silver      & 0.0173  & 0.0607  & 0.0191  & 0.0230  & 0.0219 \\
    Palladium   & \textbf{-0.0102}$^{*}$ & 0.0001  & -0.0061 & -0.0070 & -0.0089 \\
    Platinum    & \textbf{-0.0190} & 0.0068  & -0.0157 & -0.0130 & -0.0161 \\
\hline
    Chromium    & 0.1814  & \textbf{-0.0590}$^{*}$ & -0.0535 & -0.0894 & 0.2805 \\
    Lithium     & 0.3145  & \textbf{-0.0610} & 0.0250  & -0.0091 & 0.1177 \\
    Manganese   & -0.0066 & -0.1294 & -0.0347 & \textbf{-0.1355} & -0.0396 \\
    REE         & 0.7648  & 0.5124  & 0.4491  & 0.4057  & 0.4397  \\
    Silicon     & 1.1209  & 0.1591  & 1.2934  & 2.7540  & \textbf{-0.0477} \\
    Vanadium    & 1.4447  & 0.7754  & 1.6972  & 1.9673  & 0.1086 \\
\hline
\multicolumn{6}{c}{\small (\textit{b}) QLIKE loss}\\\hline
     & SV(1) & SV(2) & SV(1)-J & SV(1)-t & SV(1)-L  \\   Aluminum    & -0.2296 & -0.1235 & -0.1987 & -0.2021 & \textbf{-0.2812} \\
    Cobalt      & 0.0595        & \textbf{-0.0020} & 0.0626        & 0.0530 & 0.0690        \\
    Copper      & 0.0230        & 0.0722  & 0.1050        & 0.1312  & 0.0502        \\
    Lead        & -0.2691$^{**}$ & -0.1442$^{**}$ & -0.1976$^{**}$ & -0.1932$^{**}$ & \textbf{-0.2752}$^{**}$ \\
    Molybdenum  & 0.1108        & 0.0386  & 0.1464        & 0.0711  & \textbf{-0.0122 }\\
    Nickel      & 0.0088        & 0.0349  & 0.0249        & 0.0479  & \textbf{-0.1009} \\
    Zinc        & 0.0656        & 0.0438  & 0.1056        & 0.1174  & 0.0973        \\
\hline
    Silver      & 0.0084        & 0.3839  & 0.0267        & 0.0578  & 0.0718        \\
    Palladium   & \textbf{-0.1435} & -0.0260 & -0.0932 & -0.1089 & -0.1300 \\
    Platinum    & \textbf{-0.2071} & -0.1520 & -0.1884 & -0.1826 & -0.1669 \\
\hline
    Chromium    & 0.1194        & \textbf{-0.0304} & 0.0406        & 0.0957  & 0.1389        \\
    Lithium     & 0.4215        & -0.0426 & -0.0031       & \textbf{-0.0751} & 0.1075        \\
    Manganese   & 0.0663 & 0.2472  & 0.1248  & 0.1972  & 0.0857 \\
    REE         & 0.1500        & 0.0233  & 0.6243        & 0.6814  & 0.1215        \\
    Silicon     & -0.0998 & 0.1572  & -0.0543 & \textbf{-0.1611} & 0.3154        \\
    Vanadium    & 0.4617        & 0.1905  & 0.5087        & 0.5724  & 0.8414        \\\hline
\end{tabularx}
\caption*{\scriptsize\textit{Notes}: each entry in panel (\textit{a}) of the table shows $(RMSFE_j/RMSFE_0) - 1$, where $RMSFE_0$ is the root mean squared forecast error of the $GARCH(1,1)$ forecast, and $RMSFE_j$ corresponds to one of the SV specifications: $j = SV(1), SV(2), SV(1)-J, SV(1)-t, SV(1)-L$. In panel (\textit{b}) the loss function is the QLIKE. A negative value indicates that the $GARCH(1,1)$ forecast is less accurate than that of the corresponding SV specification. Asterisks denote the rejection of the null hypothesis of equal predictive ability in the Diebold-Mariano test, with $^{*}$ indicating significance at the 10\% level and $^{**}$ at the 5\% level.}
\end{table}

\end{document}